\documentclass[useAMS]{gGAF2e}

 \def\etal{{\it et al., }} 
 
\begin{document}
\doi{10.1080/03091920xxxxxxxxx}
 \issn{1029-0419} \issnp{0309-1929} 

\markboth{MHD turbulence for Taylor-Green vortices}
     {MHD turbulence for Taylor-Green vortices}

\title{{\textit{
The dynamics of unforced turbulence at high Reynolds number \\
for Taylor-Green vortices generalized to MHD
}}}

\author{A. Pouquet$^{1,2}$, 
\thanks{$^\ast$
Corresponding author. Email: pouquet@ucar.edu
\vspace{6pt} 
\newline\centerline{\tiny{{\em Geophysical and Astrophysical Fluid Dynamics}}}
\newline\centerline{\tiny{  ISSN 0309-1929 print/ ISSN 1029-0419 online \textcopyright 2006 Taylor \& Francis Ltd}}
\newline\centerline{\tiny{ http://www.tandf.co.uk/journals}}
\newline \centerline{\tiny{DOI:10.1080/03091920xxxxxxxxx}}}
E. Lee$^{1,3}$, M.E. Brachet$^{1,4}$, P.D. Mininni$^{1,5}$ and D. Rosenberg$^1$ \\
\vspace{6pt} 
$^1$ Turbulence Numerics Team /NCAR, P.O. Box 3000, Boulder, Colorado 80307-3000, U.S.A. \\
$^2$ Earth and Sun Systems Laboratory /NCAR, P.O. Box 3000, Boulder, Colorado 80307-3000, U.S.A. \\
$^3$ Department of Applied Physics and Applied Mathematics,
Columbia University, 500 W. 120th Street, New York NY 10027, U.S.A. \\
$^4$  \'Ecole Normale Sup\'erieure, 24 rue Lhomond, 75005 Paris, France \\
$^5$ Departamento de F\'\i sica, Facultad de Ciencias Exactas y Naturales, Universidad de Buenos Aires, Ciudad Universitaria, 1428
         Buenos Aires, Argentina
\received{received xxx}}

\maketitle


\begin{abstract}
We study decaying magnetohydrodynamics (MHD) turbulence stemming from the evolution of the Taylor-Green (TG) flow generalized recently to MHD, with equal viscosity and magnetic resistivity and up to equivalent grid resolutions of $2048^3$ points. A pseudo-spectral code is used in which the symmetries of the velocity and magnetic fields have been implemented, allowing for sizable savings in both computer time and usage of memory at a given Reynolds number. The flow is non-helical, and at initial time the kinetic and magnetic energies are taken to be equal and concentrated in the large scales. After testing the validity of the method on grids of $512^3$ points, we analyze the data on the large grids up  to Taylor Reynolds numbers of $\approx 2200$.
We find that the global temporal evolution is accelerated in MHD, compared to the corresponding neutral fluid case. We also observe an interval of time when such configurations
have quasi-constant total dissipation, time during which statistical properties are determined after averaging over of the order of two turn-over times. A weak turbulence spectrum obtains which is also given in terms of its anisotropic components.
Finally, we contrast the development of small-scale eddies with two other initial conditions for the magnetic field and briefly discuss the structures that develop, and which display a complex array of current and vorticity sheets with clear rolling-up and folding.
\end{abstract}

\section{Introduction}

Magnetic fields are present in many astrophysical and geophysical flows and are often important energetically. 
Their origin -- the dynamo problem, is ill-understood with many different behaviors observed in nature; for example, the solar magnetic field evolves in time in a somewhat regular fashion, leading to the possibility of a prediction of the amplitude and onset of the next cycle
(Wang and Sheeley, 2006), whereas the terrestrial field has an erratic temporal behavior (Valet \etal 2005). 
The dynamo effect had escaped experimental verification until recently, due to the difficulty of reaching a sufficiently high magnetic Reynolds number for dynamo action to take place within a turbulent flow at low magnetic Prandtl number as occurs in liquid metals used in the laboratory; the study of such a field as produced in the experiment yields a wealth of information close to the threshold for dynamo action (Monchaux \etal 2007). However, the high magnetic Reynolds number regime is still out of reach of the experimental approach using fluids; yet, due to its nonlinearities, the full MHD problem at high kinetic and magnetic Reynolds number as encountered in the sun, the solar-terrestrial environment or the interstellar medium for example, is highly complex with many multi-scale interactions.

 It has been shown recently (Dar \etal 2001, Alexakis \etal 2005, Debliquy \etal 2005) that such interactions are noticeably more nonlocal than in the fluid case, involving widely separated scales. Moreover, several recent {\it in situ} observations in the magnetosphere and in the solar wind show the occurence of highly energetic events due to reconnection of magnetic field lines (Hasegawa \etal 2004, Sundkvist \etal 2005, Nykyry \etal  2006, Phan \etal  2006, Retin\`o \etal  2007),
as well as rotational discontinuities (Whang, 2004).
It thus becomes important to be able to study in detail multi-scale interactions in MHD, a pre-requisite to which is to have sufficient scale separation in the fluid with high kinetic and magnetic Reynolds numbers. In three spatial dimensions, this represents a serious challenge from the numerical point of view, a challenge that will necessitate all the power that the world-wide petascale effort is going to offer, and more. There are many ways one can partially circumvent this difficulty, however; among them, modeling plays a central role.
Enforcing numerically the symmetries that a given flow may have is one such way that we shall employ in this paper. This allows for substantial savings in memory usage and in CPU time (although the CFL condition applies to the high Reynolds number that will be modeled this way). Using such highly symmetrical fields -- that are extensions to MHD of the Taylor-Green (TG) flow studied first in the context of fluid turbulence (Brachet \etal 1983), we analyze the properties of MHD turbulence for several configurations 
(see Lee {\it et al.} (2008) for more details, and below)
in a turbulent regime never explored before numerically in MHD in the incompressible case at such high resolution (see Vahala {\it et al.} (2008), and Kritsuk \etal 2009 for the case of very high resolution studies of compressible MHD turbulence). 
The next section gives a description of the initial conditions;
the temporal behavior of the insulating case (or IMTG hereafter) flow is given in \S \ref{s:temp}; spectral behavior and structures are discussed in \S \ref{s:spec}, and finally Section \S \ref{s:conclu} is the conclusion.

\section{Taylor-Green flows for MHD}\label{s:eq}

The MHD equations for an incompressible flow of constant unit density with a velocity $\bm v$ and magnetic induction $\bm b$ (in units of the Alfv\'en velocity) read:
\begin{equation}
\frac{\partial \bm v}{\partial t} + \bm v \cdot \nabla \bm v = -\nabla P + \bm j \times \bm b + \nu \Delta \bm v \ ,
\end{equation}
\begin{equation}
\frac{\partial \bm b}{\partial t} = \nabla \times (\bm v \times \bm b) + \eta \Delta \bm b \  ,
\end{equation}
together with $ \nabla \cdot \bm v = 0 = \nabla \cdot \bm b$\ ;
$P$ is the pressure and $\bm j=\nabla \times \bm b$ the current density.
In the absence of viscosity $\nu$ and resistivity $\eta$, the total energy 
$E_T=<{\textbf{v}}^2/2+\bm b^2/2>=E_V+E_M$, the total cross-correlation
$H_C=<\bm v \cdot \bm b>$
and the total magnetic helicity 
$H_M= <\textbf{a} \cdot \textbf{b}>$ (where $\bm b = \nabla \times \bm a$ is the magnetic potential) are conserved in three space dimensions. The MHD equations can take a more symmetric form using the Els\"asser variables
$\bm z ^{\pm}=\bm v \pm \bm b$ with as invariants $H_M$ and $E_{\pm}=<\bm z ^{{\pm}^2}/2>=E_T\pm H_C$. Ideal flows in MHD have been studied numerically both in the two-dimensional case (see e.g. Frisch \etal 1983) and the three-dimensional case (Kerr and Brandenburg, 1999, Lee \etal 2008) including using adaptive mesh refinement (Grauer and Marliani, 2000), but in this paper we concentrate on the dissipative case with a regular grid and using a fully parallelized pseudo-spectral code written specifically to implement the symmetries of the flow (see below). Note that no uniform magnetic field is included in these computations, and there are no forcing terms either.

The simplest Taylor-Green flow can be written (Brachet \etal 1983): 
\begin{equation}
\bm v(x, y, z) = v_0 \left[ (\sin x \cos y \cos z) \bm{\hat{e}_x} - (\cos x \sin y \cos z)\bm{\hat{e}_y} \right]  \ ; 
\nonumber  \end{equation}
the velocity component in the third direction, equal to zero initially, will grow with time.
As usual, the kinetic and magnetic Reynolds numbers are defined as 
$$
R_V=v_0L^0_V/\nu \hskip0.5truein , \hskip0.5truein R_M=v_0L^0_M/\eta \ ,
$$
with $L^0_{V,M,T}$ respectively the kinetic, magnetic and total integral scales $L^0_x= 2\pi E_x^{-1}\int [E_x(k)/k]dk$ with 
$x=V,M,T$ and $E_{x}=\int E_{x}(k)dk$ the kinetic, magnetic and total energy respectively with $E_T=E_V+E_M$.

  Similarly, one can evaluate Taylor Reynolds numbers
$R_{V,M,T}^{\lambda}$ based on the kinetic, magnetic and total Taylor scales defined respectively as:
\begin{equation}
\lambda_{V,M,T} = 2\pi \left(\frac{\int{E_{V,M,T} (k) dk}}{\int{E_{V,M,T} (k) k^2 dk}}\right)^{1/2} \ ;
\label{eq:taylor}
\end{equation}
these scales can be viewed as a measure of the curvature of the field lines.
The flow is computed in a cubic box of length $2\pi$ with minimum and maximum wavenumbers $k_{min}=1$ and $k_{max}=N/3$ respectively, with $N$ the number of points in each direction and using a standard 2/3 deliasing rule.
The Taylor-Green vortex given above can be put in correspondence with the von K\`arm\`an flow between two counter-rotating cylinders as used in several laboratory experiments, including in the case of liquid metals (Bourgoin \etal 2002).
The feasibility of dynamo action was shown numerically down to magnetic Prandtl number $P_M=\nu/\eta \approx 10^{-3}$ using the TG flow in the kinematic regime in a combination of Direct Numerical Simulations (DNS) and modeling (Ponty \etal 2005), and a dynamo was recently obtained experimentally (Monchaux \etal 2007, Bourgoin \etal 2007) using that configuration.
The TG flow has also been used numerically to study sub-critical bifurcations in the dynamo regime (see e.g., Ponty \etal 2007), the hysteresis cycle being linked with changes in the velocity field associated with the Lorentz force.

 Generalizations of the Taylor-Green flow to MHD were presented in Lee {\it et al.} (2008) where the ideal ($\nu=0=\eta$) case was studied with initially the magnetic field taken as:
\begin{eqnarray}
b_x^I &=&  b_0 \cos(x)\sin(y)\sin(z)\label{eqn:btg_I1}  \\
b_y^I &=&  b_0 \sin(x)\cos(y)\sin(z) \label{eqn:btg_I2} \\
b_z^I &=&  -2 b_0 \sin(x)\sin(y)\cos(z) ;
\label{eqn:btg_I3}\end{eqnarray}

\noindent the velocity-magnetic field global correlation $H_C\equiv 0$ for this flow and remains so at all times because of the imposed symmetries, although it has been known for a long time that there may be strong local correlations corresponding to local alignment of the velocity and magnetic field (Passot \etal 1990). 
The magnetic induction $\bm b^I$  
is everywhere perpendicular to the walls (and therefore the current $\bm j = \nabla \times \bm b$ parallel to the walls) of the so-called  impermeable box defined as $[0,\pi]^3$; the impermeable box thus appears to be insulating, and this flow is henceforth being named Insulating Magnetic Taylor-Green flow (IMTG hereafter). 

An alternate initial conditions for the magnetic field, $B_A$ hereafter, is insulating as well:
$$
b_x^{A}  = b_0^A \cos(2x)\sin(2y)\sin(2z) \  , \ 
b_y^{A} =  -b_0^A \sin(2x)\cos(2y)\sin(2z)  \ , \ 
b_z^{A} = 0 \ ,
$$
with again zero total cross helicity.
Finally, one can also construct a set of initial conditions, labeled ``conducting'' (the current being perpendicular to the impermeable box):
$$ 
b_x^{C}  = b_0^C \sin(2x)\cos(2y)\cos(2z) \ , \ 
b_y^{C}  = b_0^C \cos(2x)\sin(2y)\cos(2z) \ , \ 
b_z^{C}  = -2 b_0^C \cos(2x)\cos(2y)\sin(2z) .    
$$ 
In this latter configuration ($B_C$ hereafter), $H^C$ is non-zero but very weak (less than $4\%$ at its maximum, a dimensionless measure of correlation, relative to the total energy), whereas the magnetic helicity is zero in all three configurations.
The parameters $v_0$ and $b_0$ are chosen so that at initial time, the kinetic and magnetic energies are equal with $E_T=0.25$.

Within the periodic cube of length $2\pi$, mirror symmetries about the planes $x=0, \ x=\pi, \ y=0, \ y=\pi, \ z=0,$ and $z=\pi$ are present and will be enforced numerically,
together with rotational symmetries of angle $N\pi$ about the axes $(x,y,z)=(\frac{\pi}{2},y,\frac{\pi}{2})$ and $(x,\frac{\pi}{2},\frac{\pi}{2})$ and of angle $N\pi/2$ about the axis $(\frac{\pi}{2},\frac{\pi}{2},z)$, for $N \in Z$ (see Brachet \etal 1983, for more details). Note that the flow has several nulls (three components equal to zero):
the central point to the impermeable box $N_1=(\pi/2,\pi/2,\pi/2)$, and the planes
$P_1=(x,0,0)$, $P_2=(0,y,0)$ and $P_3=(0, 0,z)$.
Partial nulls (one or two components equal to zero) also occur; they may be as well the sites of formation of strong current sheets and thus of reconnection events.

\section{Temporal behavior of the insulating TG flow (IMTG)} \label{s:temp}

\subsection{Is a symmetric run faithful to the full dynamics?}
\begin{figure} \begin{center}
\includegraphics[width=7cm, height=50mm]{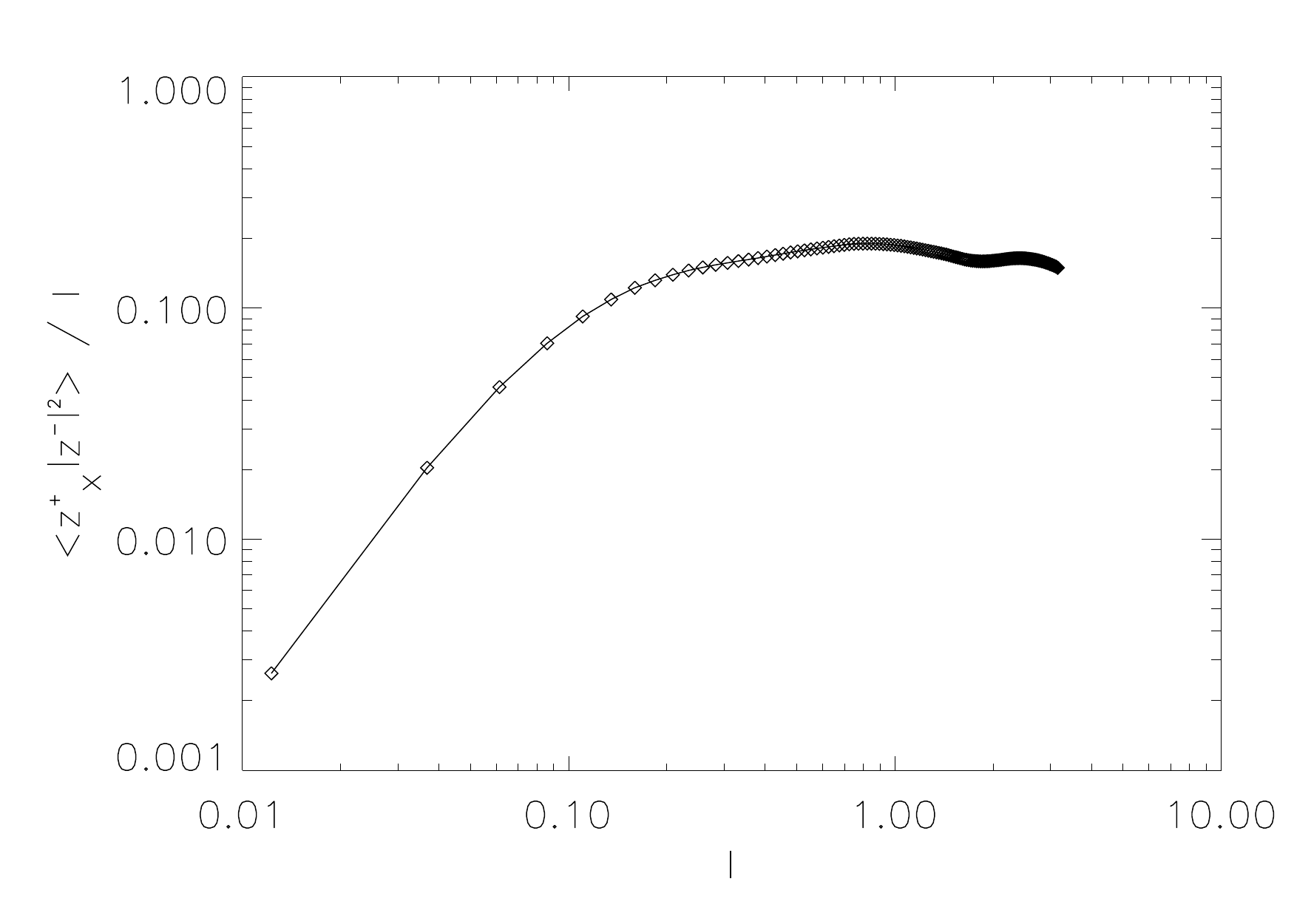} 
\includegraphics[width=7cm, height=50mm]{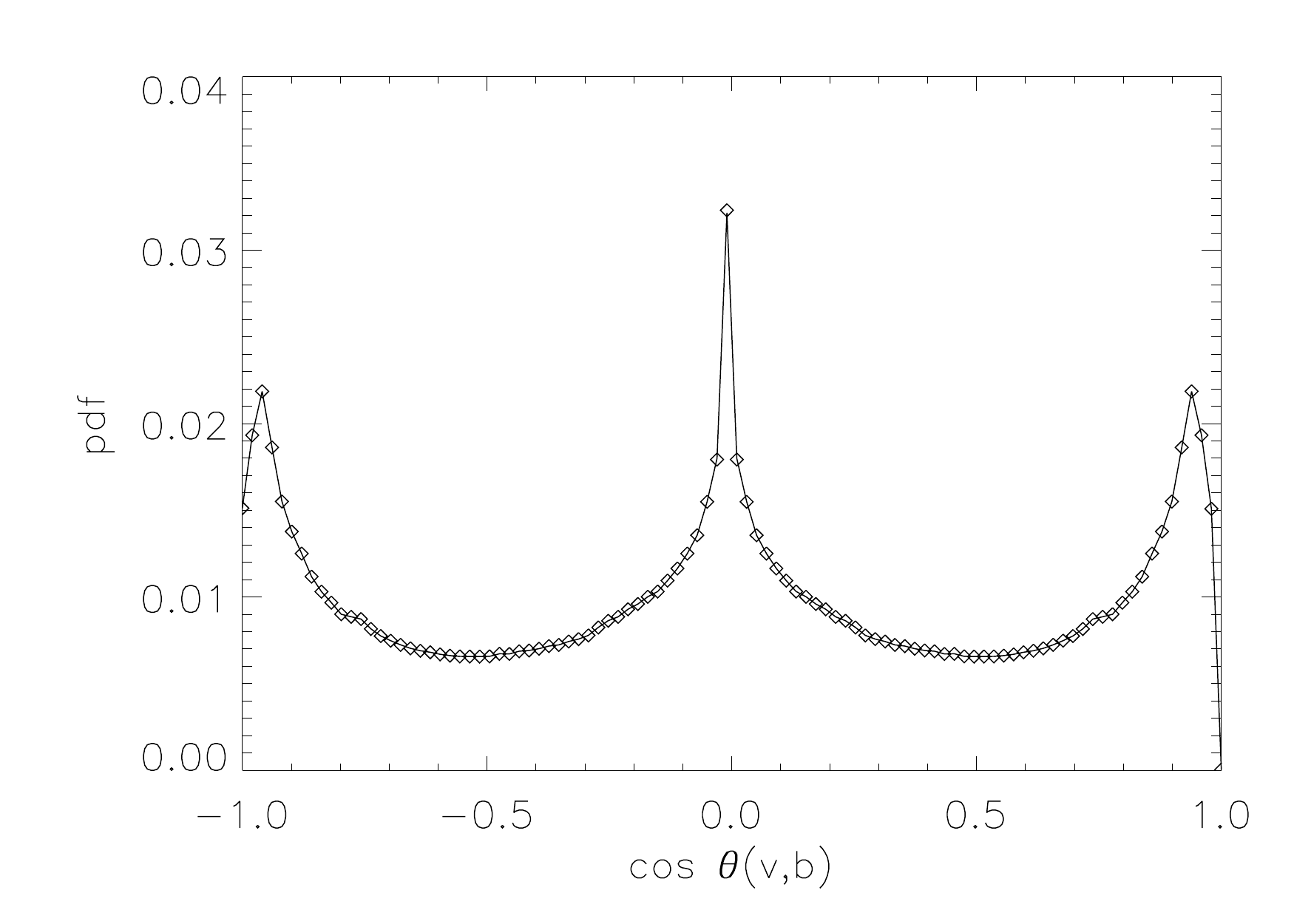}
\caption{ 
{\it Left:}  Flux-law for the ${\bf z}^+={\bf v} + {\bf b}$ variable (see eq. \ref{exactz}), compensated by $\ell$: the exact law appears as a horizontal line.

\hskip0.4truein {\it Right:} Probability distribution function of $\cos [{\bf v} \cdot {\bf b}]$. 
 Data for both figures are taken near the peak of dissipation for computations on grids of $512^3$ points, with the full run drawn with a solid line and the symmetric run with diamonds.
} \label{f_full_symm_temp1}  \end{center} \end{figure}

\begin{figure} \begin{center}
\includegraphics[width=6.cm, height=60mm]{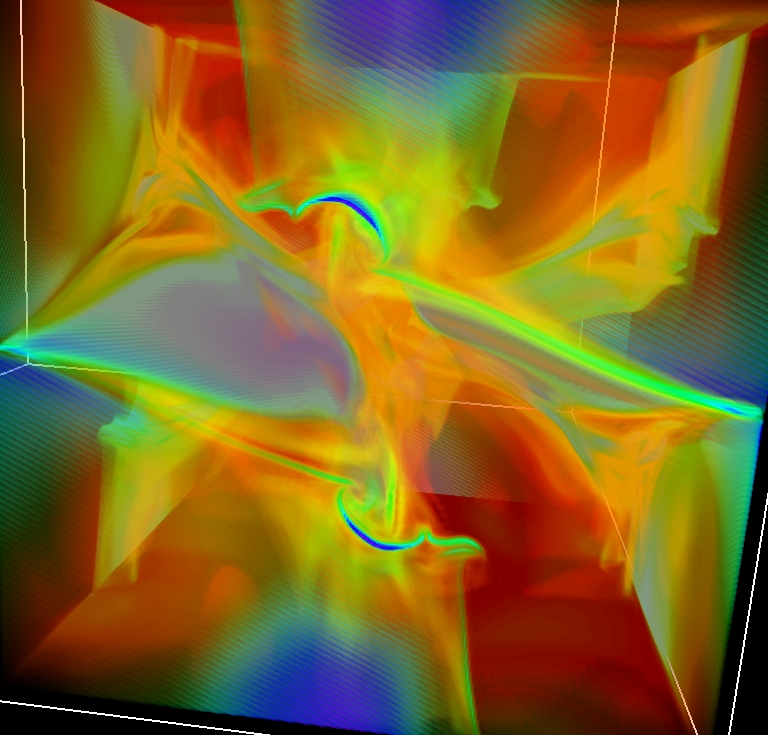}  \hskip0.5truein
\includegraphics[width=6.cm, height=60mm]{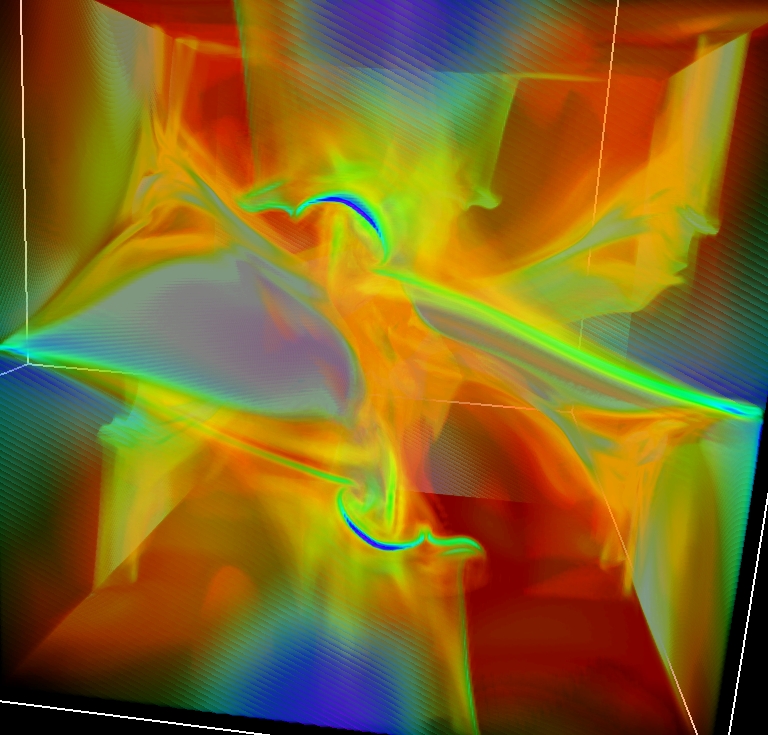} 
\caption{ 
Visualization, using the VAPOR software (Clyne \etal 2007) of the current density, on grids of $512^3$ points. 
Full DNS (right) and run implementing symmetries (left).
} \label{f_full_symm_temp2}  \end{center} \end{figure}

The numerical method we employ in this paper is based on the assumption that the implementation of the TG symmetries will not alter the overall evolution of the flow. We show in Figs. \ref{f_full_symm_temp1} and \ref{f_full_symm_temp2} that this is indeed the case, at least at the grid resolution employed for this test, using $512^3$ points. The data is taken at peak of dissipation, once the turbulence has fully developed.

The first test we performed is to verify the exact law that can be written in MHD in terms of third-order structure functions (Politano and Pouquet, 1998); in dimension three they read:

\begin{equation}
\langle \ \delta v_{L} \ (\delta v_i)^2 \ \rangle +
\langle \ \delta v_{L} \ (\delta b_i)^2 \ \rangle
-2 \langle \ \delta b_{L} \ \delta v_i \ \delta b_i \ \rangle
= -\frac{4}{3} \ \epsilon^T\ r \ ,
\label{exactv}\end{equation}
\begin{equation}
-\langle \ \delta b_{L} \ (\delta b_i)^2 \ \rangle
-\langle \ \delta b_{L} \ (\delta v_i)^2 \ \rangle
+2 \langle \ \delta v_{L} \ \delta v_i \ \delta b_i \ \rangle
= -\frac{4}{3} \ \epsilon^C\ r \ ;
\label{exactb}\end{equation}
these laws can be written in a more compact form, using the Els\"asser variables ${\bf z}^{\pm}={\bf v} \pm {\bf b}$, with associated energies $E^{\pm}$:
\begin{equation}
\langle \delta z^{\mp}_L({\bf r})  \ 
[ \delta z_i^{{\pm}}({\bf r})]^2 \rangle
=-{{4}\over{3}} \epsilon^{\pm}\ r \ ,
\label{exactz}\end{equation}
where $\epsilon^{\pm}=\epsilon^T\pm \epsilon^C$ are the energy transfer rates of $E^{\pm}$, and $\epsilon^T$ and $\epsilon^C$ the rates for total energy and velocity--magnetic field correlation. These laws, under the assumptions of incompressibility, homogeneity, isotropy, stationarity and large Reynolds number  are nothing more than the expression of the invariance of quadratic functionals of the fields under the MHD dynamical equations; as usual, $\delta f_i(r)=f_i({\bf x+r})-f_i({\bf x})$ is the difference for the i$^{th}$-component $f_i$ of the field $f$, and $f_L$ is its longitudinal component, i.e. the vector field projected on the direction ${\bf r}$ along which the difference is taken. The laws in terms of the Els\"asser fields show that the two non-helical invariants, $E^{\pm}$ or equivalently $E_T$ and $H^C$, are coupled; this leads to a double direct cascade towards small scale. In terms of the velocity and magnetic field, it shows that the conservation of $H^C$ is on equal par with that of total energy $E_T$ and that {\it a priori} two time scales can be expected to play a role in MHD dynamics, associated with the two invariants which are known to provide a partition of phase space (together with magnetic helicity), as shown in Stribling and Matthaeus (1991).
 
We see in Fig. \ref{f_full_symm_temp1} (left) that the flux law is verified on a small interval of scales, representing the inertial range at that grid resolution of $512^3$ points, and that the full and symmetric runs give identical results. Similarly, when comparing the Probability Distribution Functions (PDFs) of the alignment between the velocity and the magnetic field, as measured by the cosine of the angle between the two vectors, again no distinction can be made between the full and symmetric run: they both have a central peak (corresponding to orthogonality of ${\bf v}$ and ${\bf b}$) due to a persistence of initial conditions (not shown), and a dynamical tendency toward alignment, as observed in many MHD flows (see for a recent discussion, Matthaeus \etal 2008, Servidio \etal 2008). Finally, a three-dimensional snapshot of the current density (Fig. \ref{f_full_symm_temp2}) for the whole computational box at peak of dissipation shows that the same type of structures appear in configuration space, with a predominance of sheets, some with strong curvature.

\subsection{Energetics of the IMTG flow}

\begin{figure}[ht]  \begin{center}
\includegraphics[width=8.5cm, height=50mm]{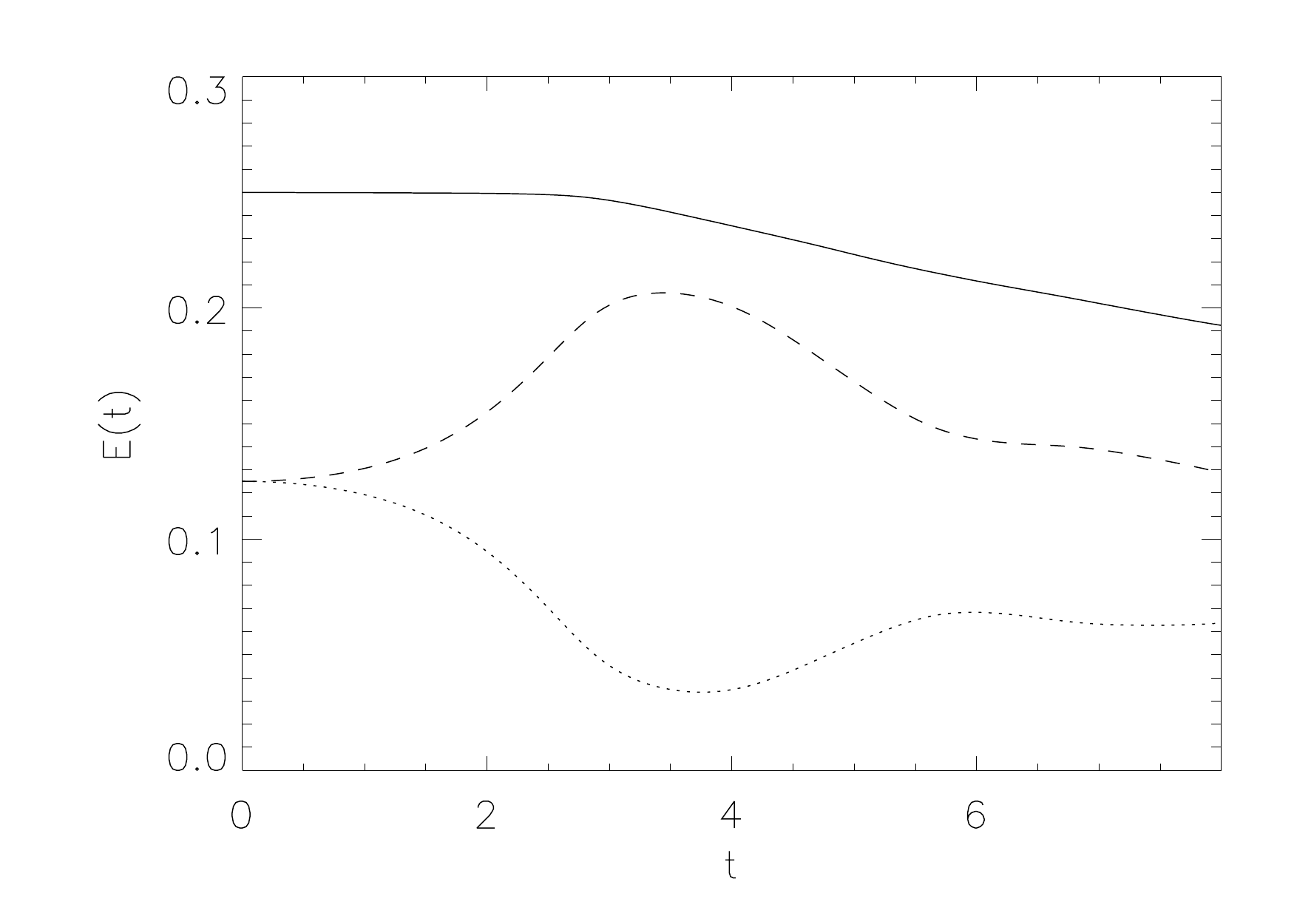} 
\includegraphics[width=8.5cm, height=50mm]{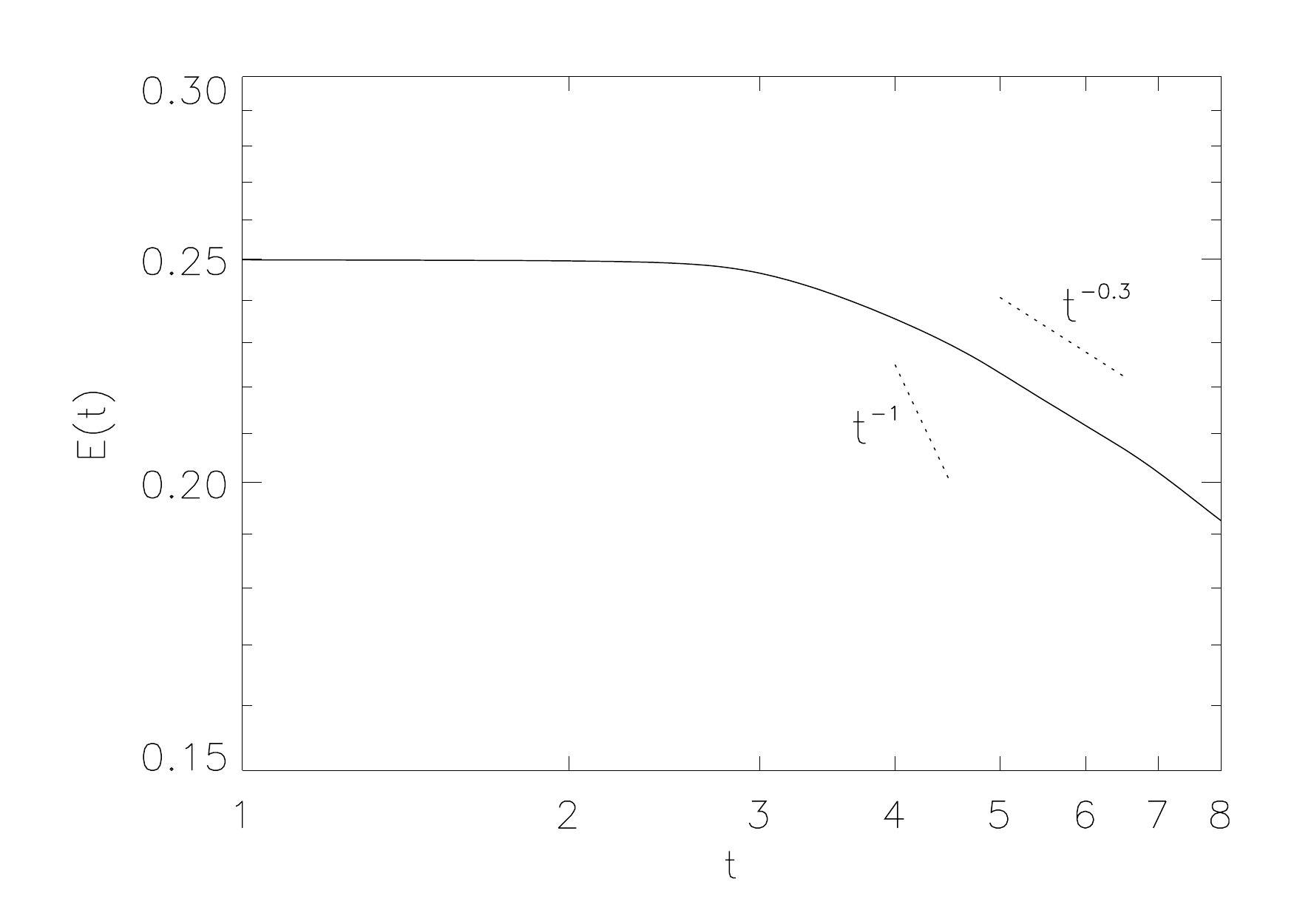}
\caption{{\it Left:} Total (solid), kinetic (dot) and magnetic (dash) energies as a function of time for the IMTG flow; equivalent grid of $2048^3$ points, Taylor Reynolds number of $R_{\lambda}\sim 2200$; note the excess of magnetic energy at all times, and the slow decay of the energy compared to the fluid case.
{\it Right:} In log-log coordinates, the total energy decay  appears to follow a weak power law.}
\label{f_temp_energy}  \end{center}
 \end{figure}

We now describe the overall energetics of the IMTG initial condition given in Eqs. (\ref{eqn:btg_I1},\ref{eqn:btg_I2},\ref{eqn:btg_I3}) in the presence of dissipation at high Reynolds number. 
 At first, the ideal behavior analyzed in Lee {\it et al.} (2008) is recovered with an exponential decay of small scales corresponding to the formation of current and vorticity sheets, and a quasi conservation of total energy, even though an exchange between the kinetic and magnetic energy $E_V$ and $E_M$ is occurring as can be seen in Fig. \ref{f_temp_energy}; this exchange can be related to an Alfv\'enic effect based on the large-scale magnetic field (recall that there is no imposed uniform field in these runs). The energy ratio $E_M/E_V$ is larger than unity at all times for this flow; a similar evolution obtains for the alternate flow $B_A$ whereas, in the case of the conducting flow $B_C$, the ratio is smaller than unity after a short transient (see Table \ref{tab1} for  more details).
 When displaying the temporal evolution in log-log coordinates for the total energy of the IMTG flow, an approximate power-law decay can be observed for a short time, with an index clearly smaller than what a Kolmogorov analysis would give: the energy decay in the Navier-Stokes case is expected to vary as $(t-t_*)^{-10/7}$ with $t_*$ a characteristic time of decay of the energy, of order unity here given our initial conditions; however, in the case of the TG fluid flow, it has been found (Brachet \etal 1983) that energy decay follows a $(t-t_*)^{-2}$ power law, a law that can be recovered phenomenologically using an argument based on the fact that the integral scale cannot grow in the TG flow used here because the initial conditions are centered on the largest available scale (Cichowlas \etal 2005).
 
Slower temporal behavior in MHD, compared to the pure fluid case, has been observed by a number of authors (Hossain \etal 1995, Kinney \etal 1995, Galtier \etal 1997, 1999, Bigot \etal 2008). It
 can be attributed to the slowing-down of energy transfer to small scales due to the interactions between waves and turbulent eddies as modeled originally by Iroshnikov (1963) and Kraichnan (1965) under the simplifying assumption of global isotropy
(see Goldreich and Sridhar, 1997,  and Ng and  Bhattacharjee, 1997, for a straightforward extension of that phenomenology to the anisotropic case). This slowing-down can be understood when writing the MHD equations in terms of the Els\"asser variables ${\bf z}^{\pm}={\bf v}\pm {\bf b}$: the nonlinearities involve the products ${\bf z}^{\pm} \cdot \nabla {\bf z}^{\mp}$ but there are no self interactions ($z^+z^+ $ or $z^-z^-$). In MHD, following the same phenomenology as Kolmogorov but taking into account the effect of Alfv\'en waves, the temporal decay of energy can be shown to become, in three dimensions 
$\sim (t-t_*)^{-5/6}$ under such assumptions. Note that a further slowing-down of energy decay may stem from the presence of a strong large-scale field, as shown recently in the context of anisotropic MHD (Bigot {\it et al.}, 2008), with now a decay $\sim (t-t_*)^{-2/3}$. 
 However, the slowing-down observed here, with $E(t)\sim (t_*-t)^{-0.3}$, is even more important. 
 The origin of this behavior is not known. It may be related to the strong (relative) increase of magnetic energy, with $E_M/E_V$ reaching a maximum $\sim 4 $ at $t=2.5$ and decreasing steadily thereafter to a value close to $1.5$ at the end of the run; this may entail a faster decay at later times that could only be observed at higher Reynolds numbers, although this peak in the magnetic to kinetic energy ratio augments monotonically with Reynolds number, from $E_M/E_V\sim 2.5$ for $R^{\lambda}\sim 120$ (see also Table \ref{tab1}).
 This slow decay could also be due to the local alignment between the velocity and magnetic field although the total correlation remains weak in relative terms (less than 4\%). Tendency toward alignment of ${\bf v}$ and ${\bf b}$ is obtained in many observational and numerical MHD flows  as mentioned earlier (Matthaeus \etal 2008, and references therein); it weakens the nonlinear terms responsible for the decay of energy at high Reynolds number; indeed, 
slow decay can be observed in the presence of strong and global correlations (Galtier \etal 1999). Another feature specific to the IMTG flow is that, at $t=0$, the magnetic field and the vorticity are identical (and anti-parallel). Such is not the case for the other two flows studied in this paper, for which a faster decay obtains with the total energy varying as $(t-t_*)^{-1.05}$, although still slower than for the pure fluid TG case.
Obviously higher resolutions would be needed to clarify this point for the IMTG flow since it would allow for a larger temporal range in the self-similar decay of the energy 
before the energy and thus the Reynolds numbers drop too significantly.

 \begin{figure}[ht] \begin{center}
\includegraphics[width=8.5cm, height=50mm]{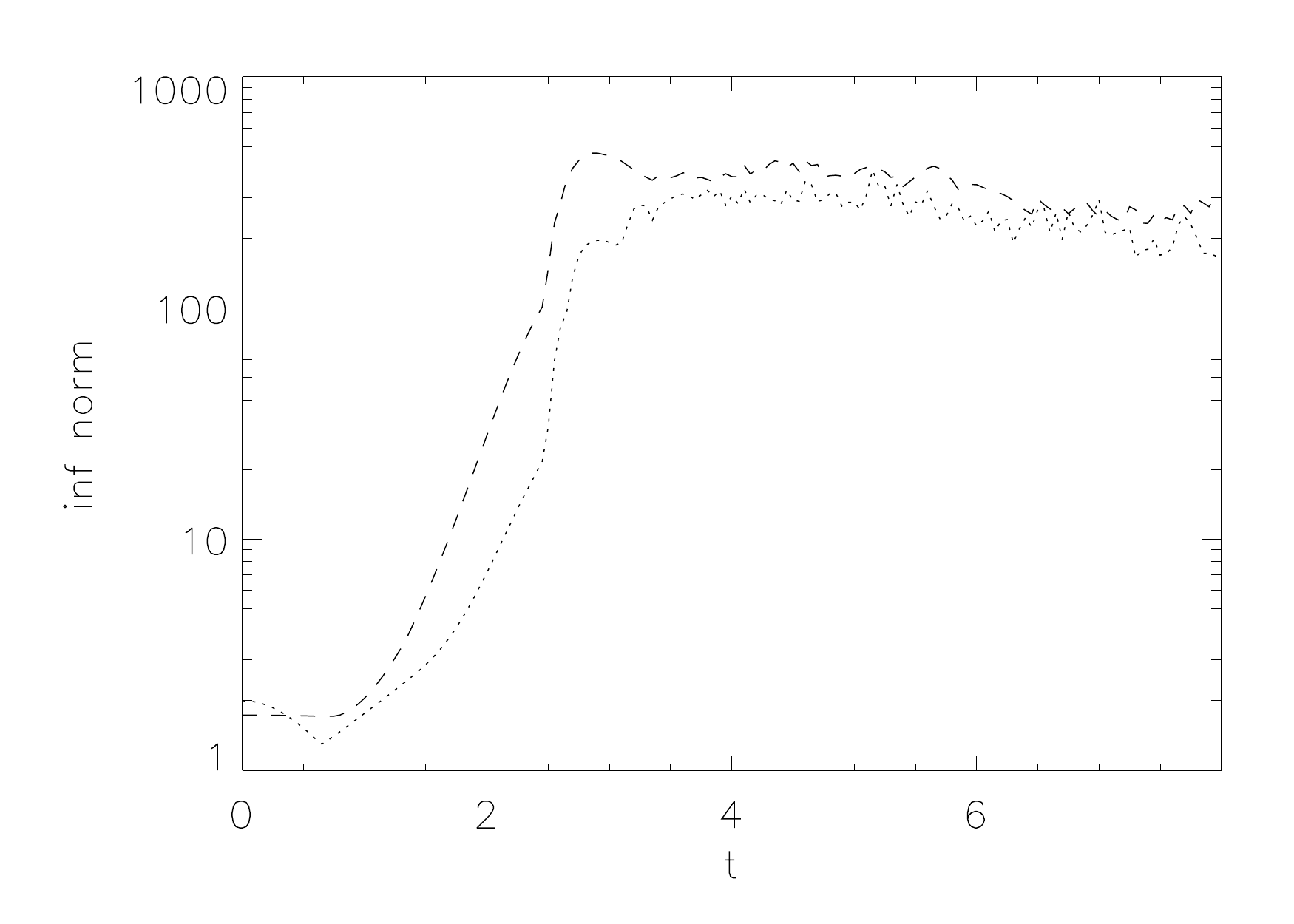}
\includegraphics[width=8.5cm, height=50mm]{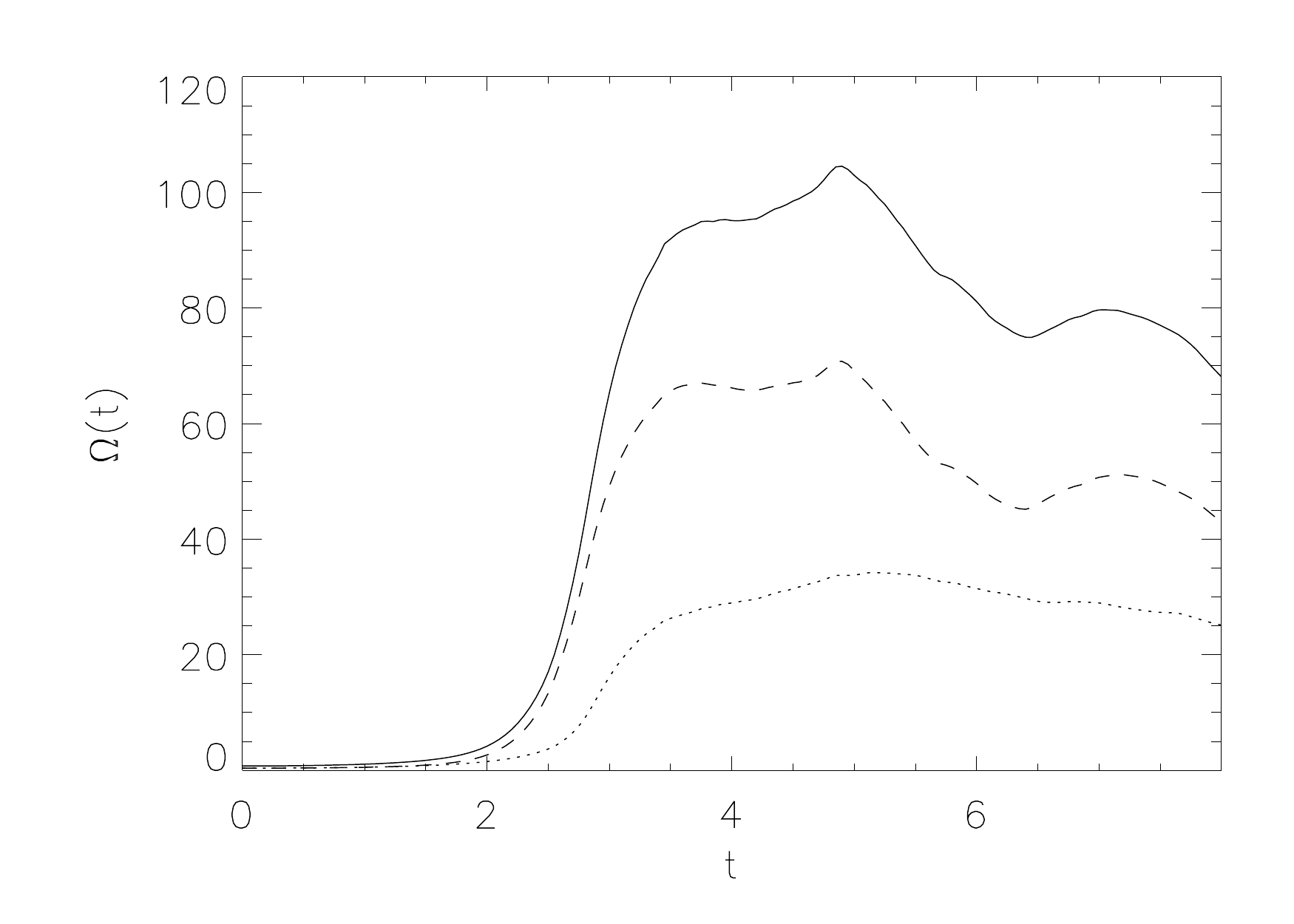} 
\caption{
{\it Left:} Maxima of the current (dash line) and vorticity (dotted line) as a function of time, with a slight excess of the former.

\hskip0.43truein  {\it Right:} Temporal evolution of dissipation (kinetic, dotted line,  magnetic, dash line, and total, solid line) as a function of time. 

\hskip0.33truein Note the plateau first reached around $t\approx 3.5$ and lasting for a couple of eddy turn-over times, and subsequent peaks likely associated with reconnection events of current sheets. 
} \label{f_temp_enstrophy} \end{center} \end{figure}

When examining the temporal evolution of the maxima of the current (dash line) and the vorticity (dotted line), given in Fig.  \ref{f_temp_enstrophy} (left), we observe that the first initial exponential phase is ideal and corresponds to thinning of current and vorticity sheets due to large-scale shear; this phase is interrupted rather abruptly at $t\approx 2.5$ by another phenomenon that now evolves more rapidly and corresponds to the quasi-collision of two current sheets, pushed together by the contribution  of the Lorentz force to the magnetic pressure on either sides of the sheets (see also Fig. \ref{f_temp_skew2} in the next Section). This is still in the ideal non-dissipative phase and gives rise to a quasi rotational discontinuity (Lee \etal 2008) which is then arrested, around $t\approx 2.65$, by dissipation setting in at the smallest scales. The dissipative phase in the {\it sup} norms can be seen as somewhat chaotic but a characteristic feature emerges, that is that the ratio of $j_{max}/\omega_{max}$ remains approximately constant, including at the latest time of the computation when roughly 28\% of the energy has already been lost to dissipation. This is somewhat reminiscent of the constancy of the cancellation exponent $\kappa$ in a turbulent flow (Graham \etal 2005), where $\kappa$ measures by how much the change in sign of (say) the magnetic field at a given scale varies with scale; it can be attributed to the fact that in the self-similar decay of energy, as long as the Reynolds number remains sufficiently large, the complexity of the flow (as measured for example by $\kappa$) remains approximately the same even though the energy itself does decay, at an algebraic slow rate.

In Fig. \ref{f_temp_enstrophy} (right) is shown as a function of time the total generalized enstrophy $\Omega^T=\large< \omega^2/2 \large> + \large< j^2/2 \large>$ (solid line), proportional to energy dissipation since $\nu=\eta$,
as well as its kinetic (dots) and magnetic (dash) components. After an initial exponential growth, the IMTG flow displays a sizable plateau during which the dissipation of energy remains quasi constant and thus during which temporal averages can be performed in order to study with greater accuracy the statistics of the flow such as its spectral behavior (see below). Again, the small scales are dominated by the magnetic structures; also note the presence of later maxima at $t\approx 5$ and $t\approx 7$, probably a sign of renewed reconnection events of magnetic field lines; the second peak around $t=5$ appears at this resolution but is barely perceptible at lower Reynolds numbers.

 \begin{figure}[ht] \begin{center}
 \includegraphics[width=8.5cm, height=50mm]{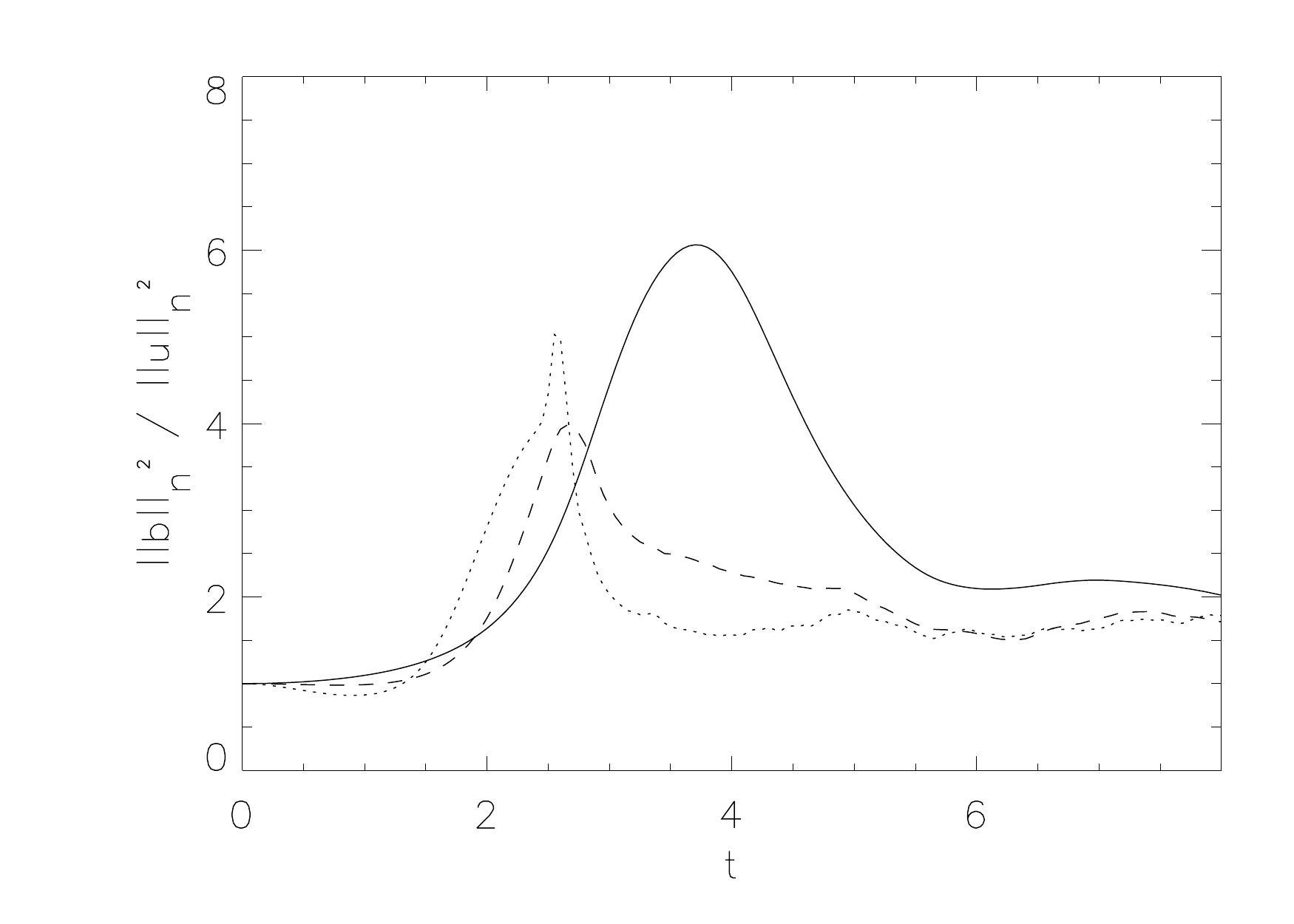}
\includegraphics[width=8.5cm, height=50mm]{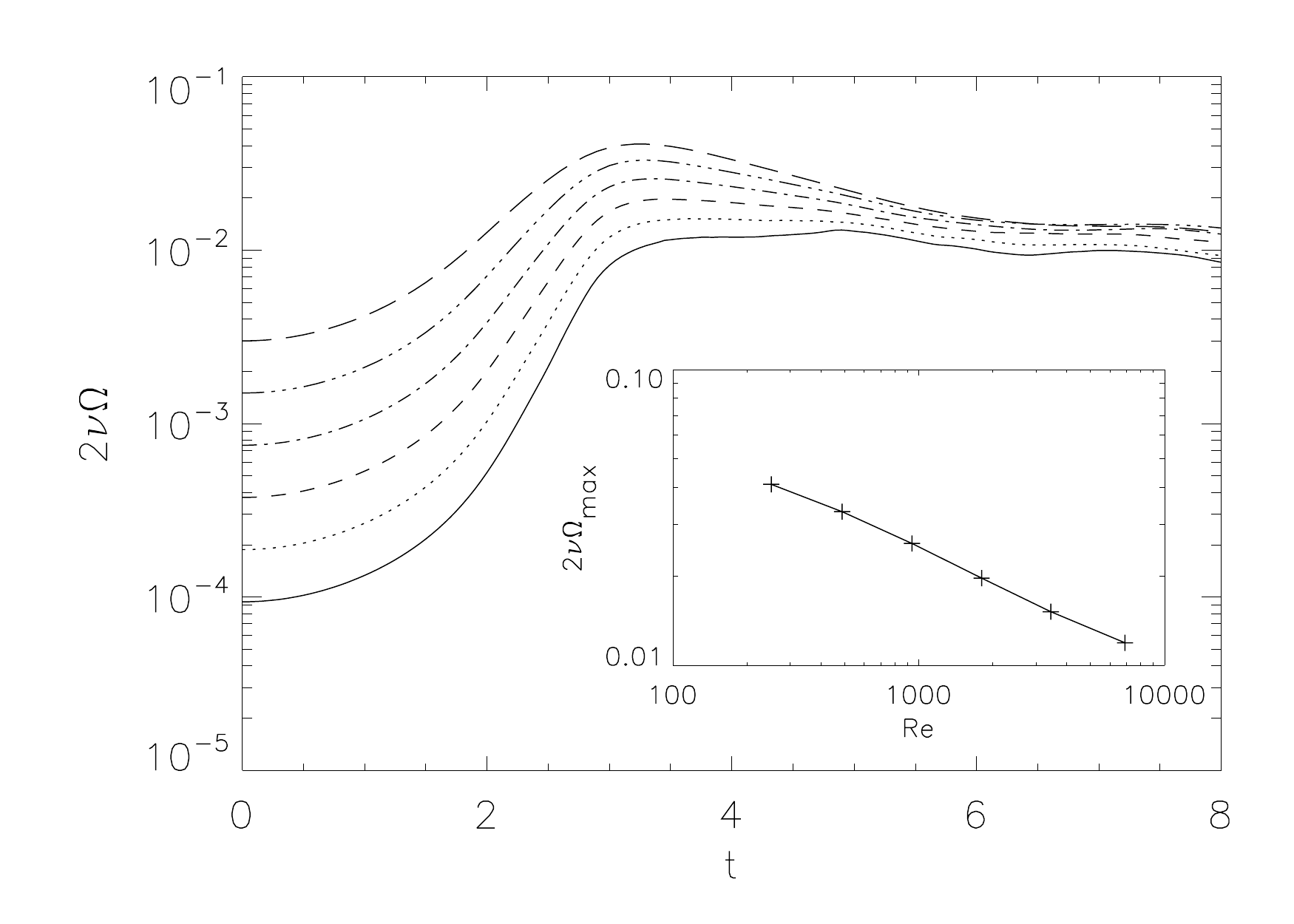}
\caption{
{\it Left:} Temporal variation of the magnetic to kinetic ratios of the first three moments of the basic fields (see Eq. \ref{sob}) with $n=0,1,2$ represented with a solid, dash and dotted line respectively. All display an excess of magnetic excitation, with an earlier peak for those moments involving derivatives and corresponding to an abrupt acceleration in the development of small scales in that flow. 

{\it Right:} Temporal variation as a function of Reynolds number (and thus grid resolution) of total dissipation for the IMTG flow, from a grid of $128^3$ points (long dash) to $2048^3$ (dots) in lin-log coordinates; the inset in log-log coordinates gives the first peak of dissipation as a function of $R_V$; note the tendency for it to slowly decrease with Reynolds number for the IMTG flow.
} \label{f_enstrophy_R} \end{center} \end{figure}

This is corroborated by the examination of the ratio of  magnetic to kinetic excitation for several moments of the fields displayed in Fig. \ref{f_enstrophy_R} (left); these ratios are defined as:
\begin{equation}
R_n = \frac{\int k^{2n}E_M(k)dk } {\int k^{2n} E_V(k)dk} \ ,
\label{sob} \end{equation}
and can be related to the respective Sobolev norms. There is a systematic overshoot of magnetic excitation, a well-known feature of turbulent flows for example often observed in the solar wind for the energy ratio $R_0$; it may possibly be due to the fact that turbulent magnetic resistivity is known to be less efficient than turbulent viscosity (Pouquet \etal 1976), a phenomenon modeled through the use of second-order closures of turbulence. Small--scale structures are dominated as well by magnetic excitation, to a somewhat lesser extent as $n$ increases and with an earlier peak at $t\approx 2.5$ when the first strong current structure has formed (see Lee \etal 2008, and Fig. \ref{f_temp_skew2}). Furthermore, $R_n\rightarrow 1$ as $n$ increases (and also as time elapses); this could be interpreted as being due to the fact that, as we increase the weight of the small scales, these global averages get to be dominated by a few localized events in space that tend to be similar in their physical structure. 

The variation of total dissipation with time of the IMTG flow is given in Fig. \ref{f_enstrophy_R} (right) in lin-log coordinates for several Reynolds numbers. We observe that the initial phase shows less dissipation when the viscosity is reduced (from $2\times 10^{-3}$ (long dash) to $6.25 \times 10^{-5}$ (solid line), with runs on grids of $128^3$ points to $2048^3$ points, augmenting by a factor of $2^8$ and with two different runs at the largest resolution. The inset gives, in log-log coordinates, the variation with Reynolds number of the first temporal maximum of dissipation at the end of the ideal phase; for this flow, there seems to be a steady (power-law) decrease of dissipation, although constant energy dissipation is observed in MHD turbulence with the $B_{A,C}$ flows, as well as in a full numerical simulations using a Beltrami flow (Mininni and Pouquet, 2007). 
The question of finite dissipation in the limit of infinite Reynolds number in MHD will thus need further study at higher Reynolds number; indeed, one difference in behavior between the IMTG and  $B_{A,C}$ flows is the afore mentioned observation of a very slow decay of energy for IMTG when contrasted to the other flows.

\section{Spectra and structures for the IMTG flow} \label{s:spec}

\subsection{Energy spectra}

\begin{figure}[ht] \begin{center}
\includegraphics[width=8.5cm, height=50mm]{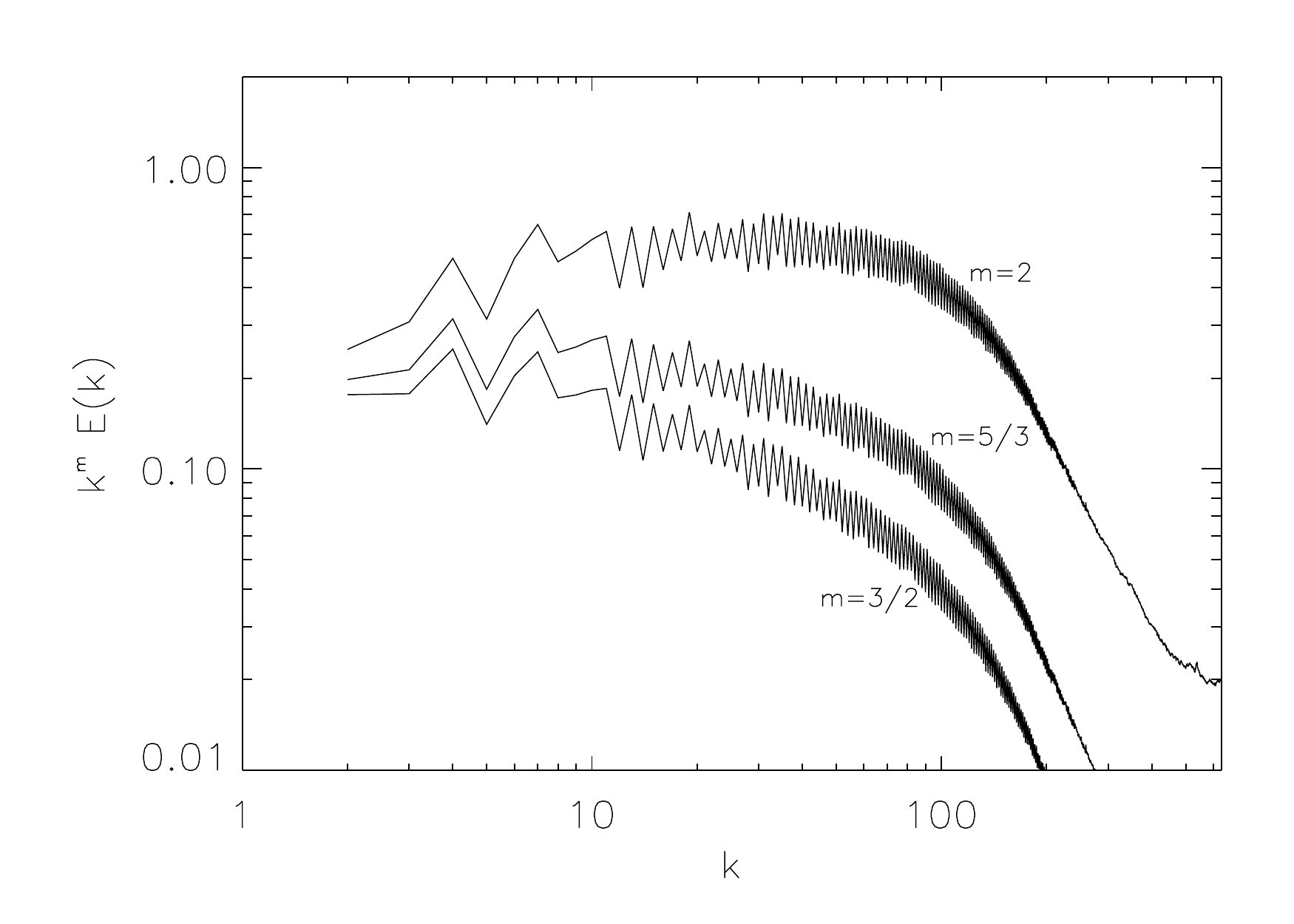}
\includegraphics[width=8.5cm, height=50mm]{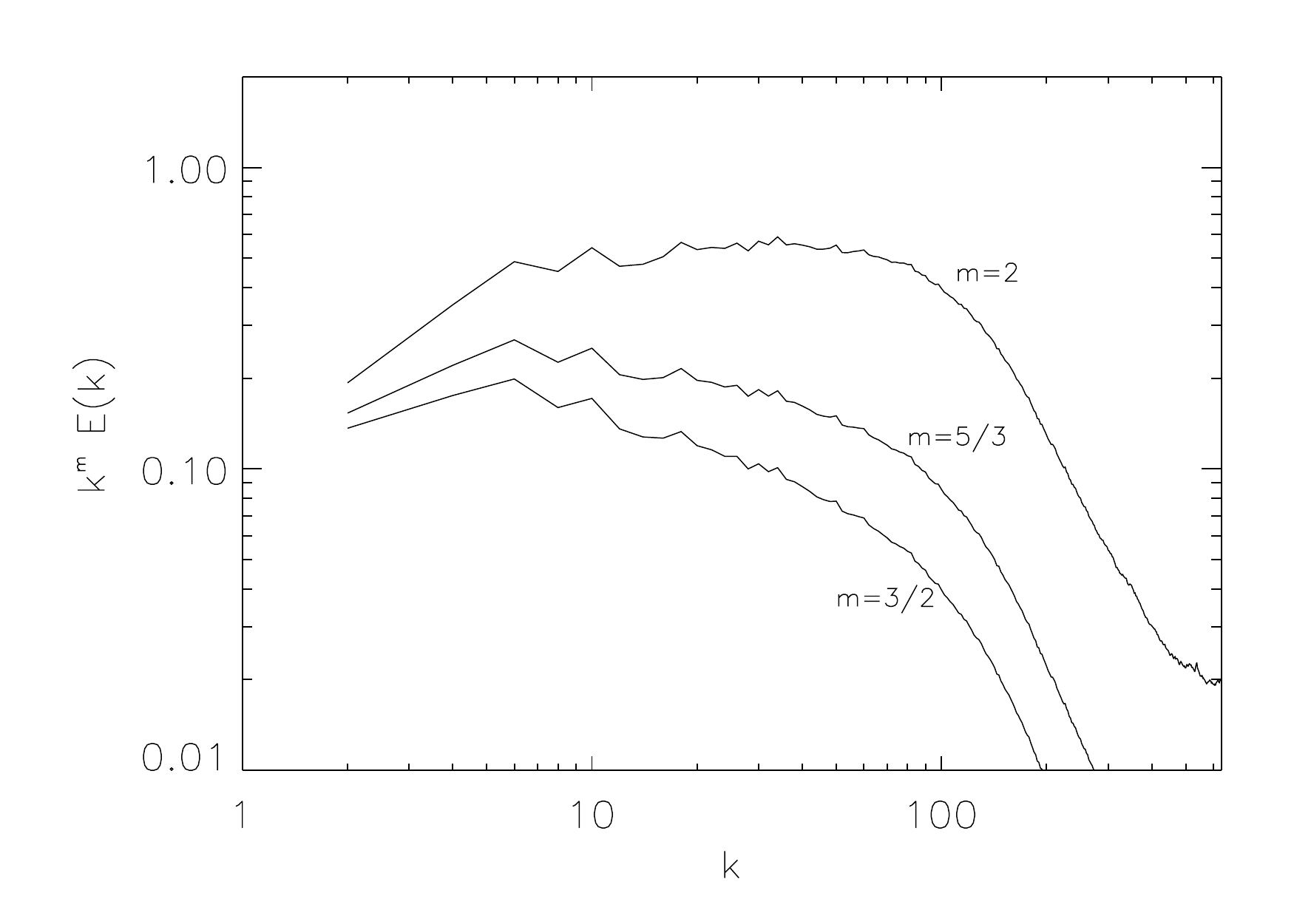} 
\caption{Total energy spectra compensated by $k^m$, with $m=2,\ 5/3$ or $3/2$ (see labels)  for the IMTG flow.
{\it Left:} instantaneous total energy near the maximum of dissipation.
{\it Right:} total energy spectra averaged in the plateau of dissipation, $t\in [3.5,5]$.
Note that in the figure at right, the spectra are also averaged between adjacent shells in order to get rid of the even-odd oscillations due to the structure of the Taylor-Green configuration. A clear $k^{-2}$ spectrum emerges from this analysis for the total energy.}
\label{f_spectra} \end{center} \end{figure}

We show in Fig. \ref{f_spectra} (left) the total isotropic energy spectrum, compensated by $k^2$ (top curve), $k^{5/3}$ (middle curve) and  $k^{3/2}$ (bottom curve), computed at $t=4$ close to the first peak of dissipation; these compensations correspond respectively to a weak turbulence spectrum (WT hereafter, Galtier \etal 2000, 2002, 2005), to a Kolmogorov (1941) spectrum (hereafter, K41) and to an Iroshnikov (1963) -- Kraichnan (1965) spectrum (hereafter, IK). Issues relating to anisotropy are discussed further below. The spectrum is instantaneous and it is computed for the run on a grid corresponding to $2048^3$ points.
Although even at such a resolution, it may be hard to discern spectral laws, the WT spectrum seems to be fulfilled better.
Fig. \ref{f_spectra} (right) is the same as Fig. \ref{f_spectra} (left) but now the data is averaged over the plateau of energy dissipation discussed previously and over adjacent shells; a $k^{-2}$ law emerges clearly when temporal averaging is employed.
Note that the kinetic and magnetic integral scales for the IMTG flow I6 (see Table \ref{tab1}) are respectively 1.61 and 1.95, the Taylor scales are 0.32 and 0.42 and the Kolmogorov dissipation scales 0.03 and 0.02.
These values obtain from a temporal averaging between t=4.5 and t=5, over 21 snapshots; we thus conclude that the flow is well resolved since the dissipation scale is measurably larger than the grid spacing, and that the inertial index for the total energy is a nonlinear effect, followed by a sizable dissipation range.
Further, we note that the Alfv\'en time is 3.3 times shorter than the eddy turn-over time at the integral scale, and is almost 14 times smaller at the Taylor scale, consistent with the fact that the spectra show no K41 behavior, the transfer time of energy through nonlinear mode coupling being affected by Alfv\'en wave propagation.

\begin{figure}[ht] \begin{center}
 \includegraphics[width=12.9cm, height=50mm]{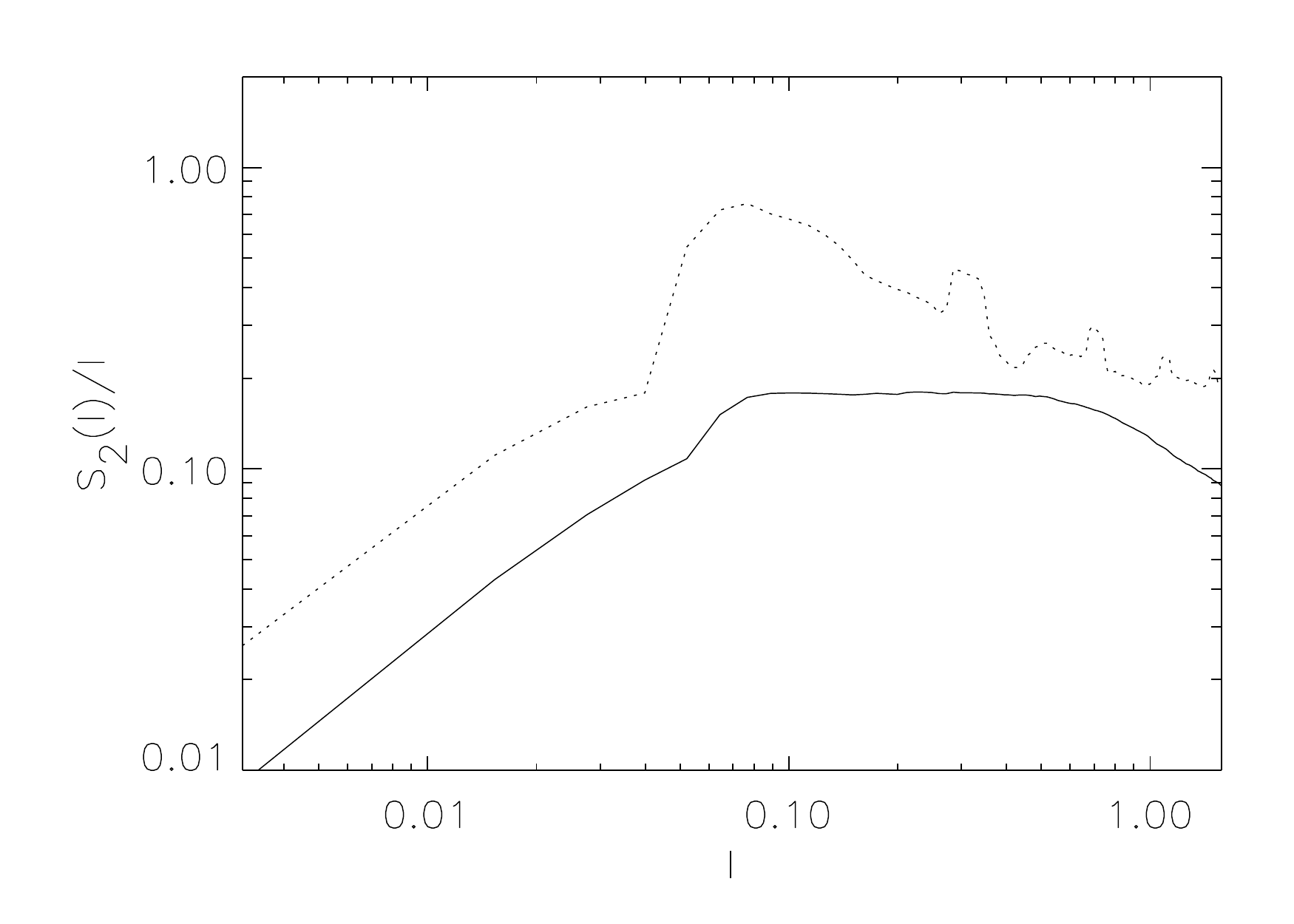} 
\caption{Perpendicular (solid line) and parallel (dotted line) components of the second-order structure function of the magnetic field, compensated both by $\ell$;
$\perp$ and $\parallel$ refer here, in the absence of an imposed magnetic field, to the direction of a locally-averaged magnetic field, in a sphere of diameter the magnetic integral scale.
Note that the perpendicular structure function follows a law $S_{2,\perp}\sim l$ over roughly a decade in scales, corresponding to a weak turbulence regime (Galtier \etal 2000). 
Similar results obtain for the velocity.}
\label{sp_aniso} \end{center} \end{figure}

In the case of weak turbulence, it has been known for a long time that the flow becomes anisotropic under the influence of a large-scale (uniform) magnetic field $B_0$; a phenomenological description of the influence a large scale magnetic field can have on the anisotropic dependence of spectra (Goldreich and Sridhar, 1997, Ng and Bhattacharjee, 1997) yields, by a simple generalization of the IK reasoning, to a $k_{\perp}^{-2}$ spectrum, as confirmed by theoretical developments (Galtier {\it et al.}, 2000). 

ADD HERE:  Weak turbulence has been observed in the magnetosphere of Jupiter (Saur \etal 2002) and more recently in a large numerical simulation (Mininni and Pouquet, 2007).
Numerous studies of two-dimensional MHD turbulence, which can be regarded as a simple model of MHD turbulence in the presence of a strong magnetic field, have been performed, including in its intermittency properties (see e.g. Sorriso \etal 2000); the extension of this work to the study of non gaussianity of weak MHD turbulence is left for future work (Lee \etal 2009).

In the absence of a uniform field, the local mean field at large scale can play an equivalent role as far as the small scales are concerned, provided there is sufficient scale separation between the two, leading to a strong energetic difference between the largest and smallest scales, by several order of magnitudes. High-resolution runs provide, to some extent, such a scale separation and we now analyze the flow in these terms, following the procedure developed previously (Mininni and Pouquet, 2007) and applying it to the magnetic energy spectrum: specifically, one averages the magnetic field in a sphere of diameter the magnetic integral scale, and defines, locally, the perpendicular and parallel dependence of structure functions with respect to that locally averaged field. The result can be found in Fig. \ref{sp_aniso} with the second-order structure function of the magnetic field displayed in terms of its perpendicular (solid line) and parallel (dotted line) components, and both compensated by $\ell$ (i.e., corresponding to a $k_{\perp,\parallel}^{-2}$ spectrum); these functions are plotted at $t=5$, at the second peak of dissipation (see Fig. \ref{f_temp_enstrophy}).
A convincing weak turbulence scaling again appears in the large scales for the energy, in terms of $l_{\perp}$, starting roughly at the magnetic Taylor scale 
($\lambda_M\sim 0.42$), whereas the smallest scales follow a regular behavior as given by a Taylor expansion, again a testimony of the fact that the flow is well resolved. The structure function in terms of its variation with $l_{\parallel}$, regular again in the smallest scales, has a more chaotic behavior in the large scales and does not seem to follow a clear power law. It should be noted that the theory for weak MHD turbulence does not give any dependence on $\ell_{\parallel}$, which thus may arise from higher-order developments. Considering the phenomenology proposed by Goldreich and Sridhar (1995), of a scale-independent equilibrium between the timescales for nonlinear eddy turn-over time and the Alfv\'en time, is a further step in the description of such flows, a point that will be studied in the future (Lee \etal 2009).

Several other features in these plots are noticeable. First of all, the magnetic field is dominated energetically at all scales by its parallel component, a feature likely dependent on both the initial conditions and the intrinsic dynamics of the flow. 
Furthermore, the ratio of the parallel to perpendicular structure function is somewhat constant at large scale, and thus it appears that there is similar transfer at large scales (scales larger than the magnetic Taylor scale) in the perpendicular and parallel components of the energy spectrum; this confirms the finding of Mininni and Pouquet (2007) where, due to initial conditions that are strongly isotropic, the large scale spectrum remains isotropic down to the magnetic Taylor scale, and only becomes compatible with weak turbulence at scales smaller than $\lambda_M$.

\subsection{Small-scale structures}
The abrupt transition in the scaling behavior of the second-order structure function of the magnetic field just discussed takes place close to the dissipation scale; this can be viewed as being indicative of the formation of sharp structures in configuration space. We give in Fig. 8
a perspective volume rendering, zooming on different structures of the IMTG flow at time $t=5$. The opacity for the plots is such that only features above a few r.m.s. value are displayed.
On the left is a current sheet which has formed a roll; the blue and purple lines indicate magnetic field lines which display the formation of a cusp-like feature close to the core of the rolling current sheet; this roll is roughly $\pi/20$ across, which seems typical when one examines other current sheets in the computational box, with some variation in transverse size and with a length of the order of the magnetic integral scale.
In Fig. 8 (middle), the vorticity is plotted at the same time and at the same location in space. This confirms earlier findings (Mininni and Pouquet, 2007) of strong correlation between the velocity and magnetic field in the small scales (as characterized by the behavior of vorticity and current), even though the global correlation remains small (of the order of a few \% as mentioned before); this implies that, with weak nonlinear terms in such structures, they will survive for a time of the order of an eddy turnover time, similar to the case of a fluid within which the vortex filaments display a strong correlation between velocity and vorticity. Note that the curl of the Els\"asser variables, $\omega^{\pm}=\omega\pm j$, are also co-located, with a dominance of $|\omega^-|$ by 36\%.

\begin{figure}[ht]  \begin{center}
      \hskip-0.39truein                            \includegraphics[width=7.cm, height=60mm]{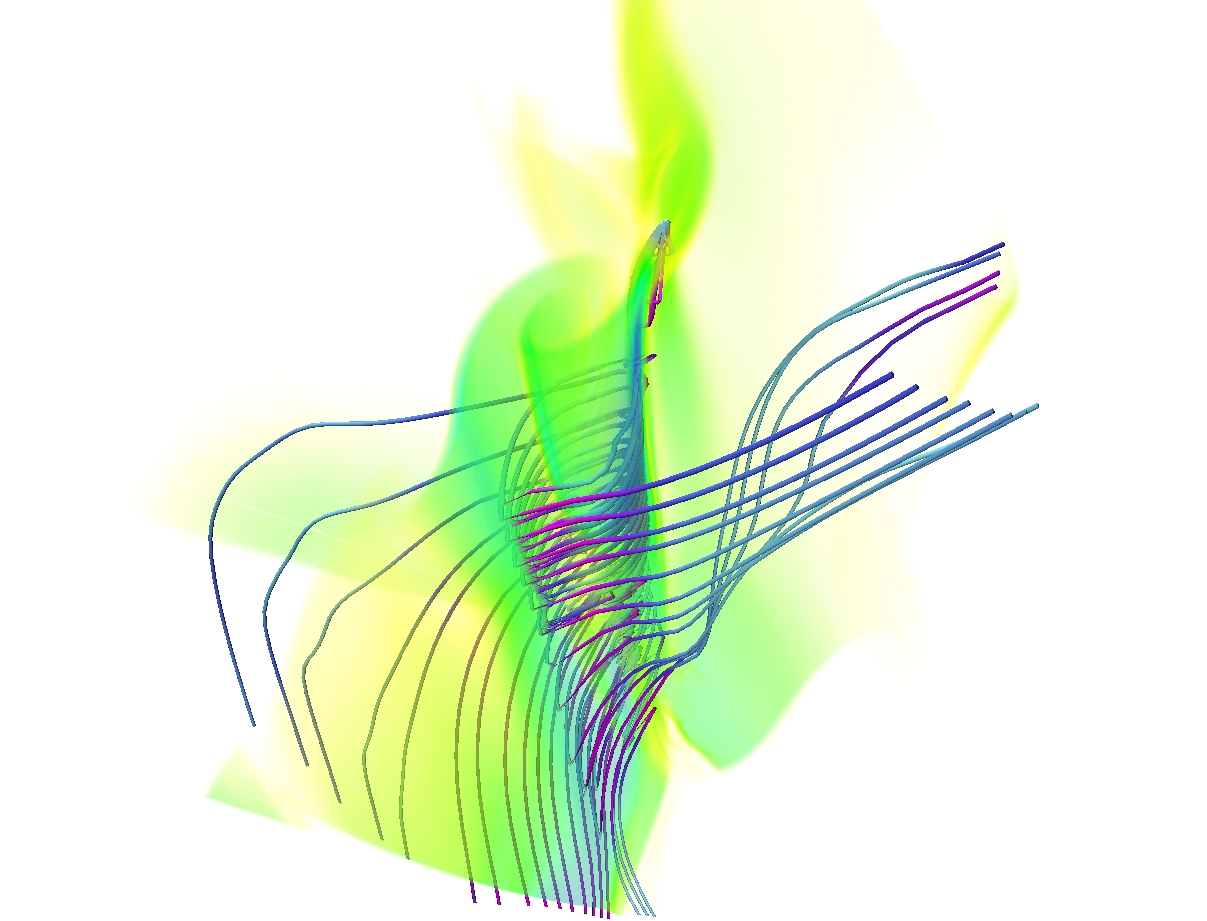}  \hskip-0.60truein
                                                                \includegraphics[width=7.cm, height=60mm]{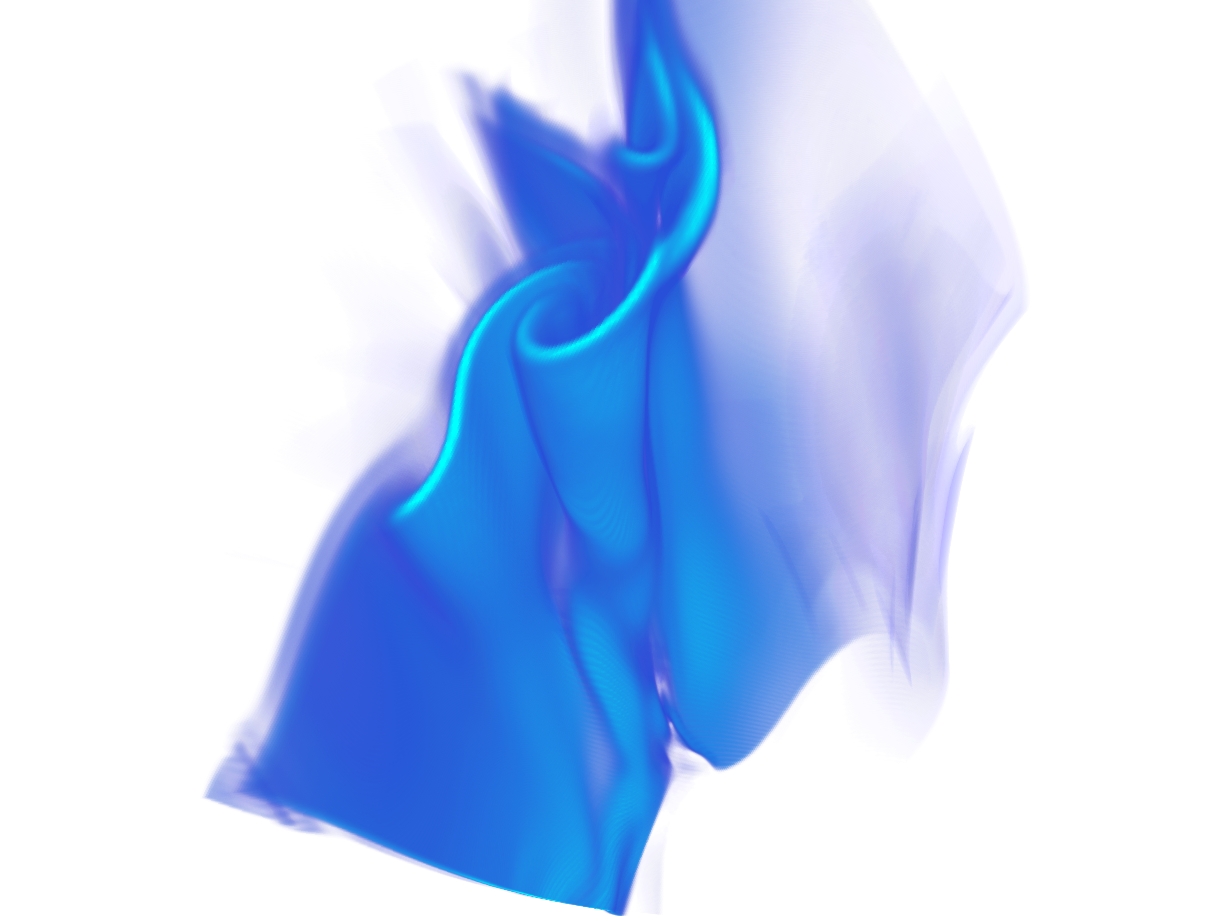}  \hskip-0.40truein
                                                                \includegraphics[width= 7.cm, height=60mm]{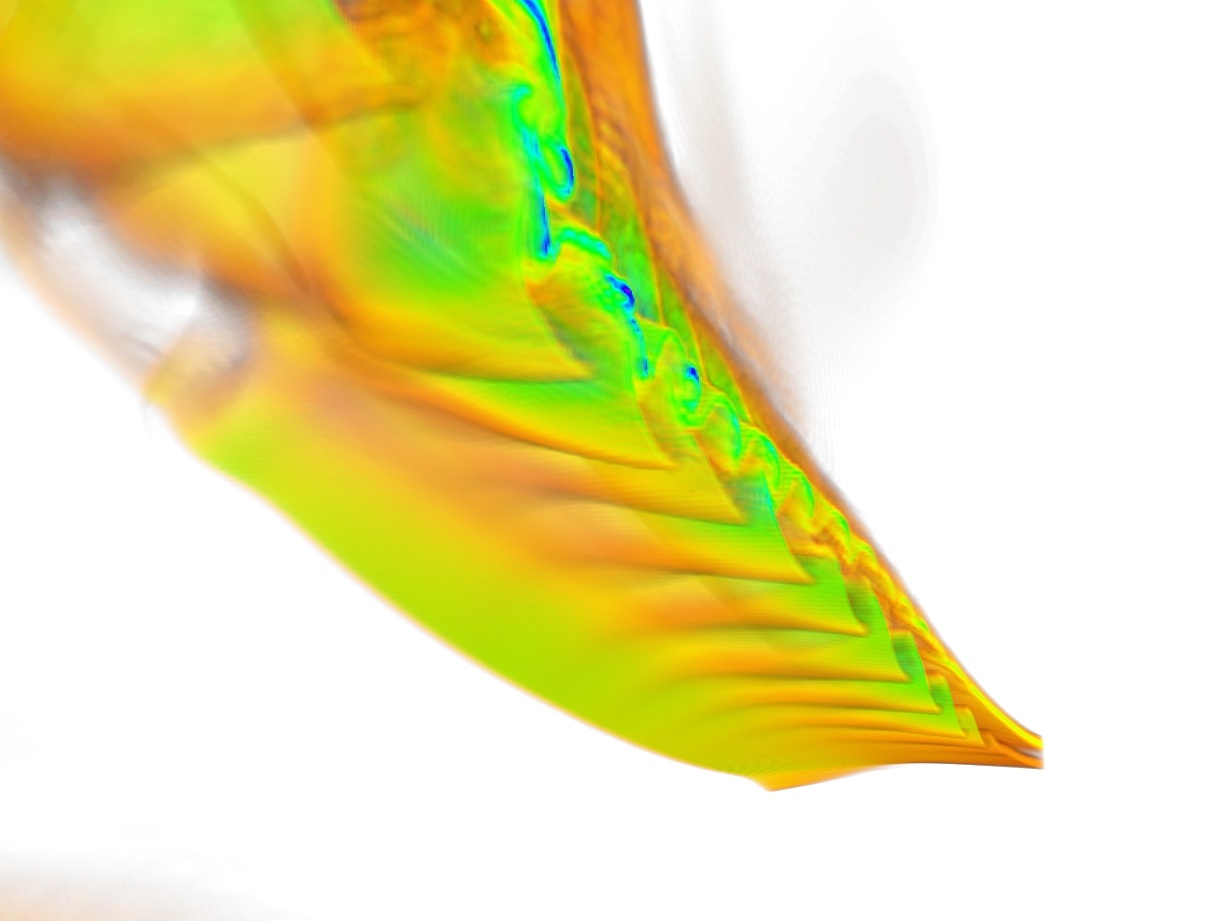}
\caption{Perspective volume rendering of a zoom on the intensity of the current (left), the vorticity (center) and another sub-volume for the current (right), all taken near peak of dissipation for the  IMTG flow on a grid of $2048^3$ points, with a Taylor Reynolds number of $\sim 2200$. Note the curved vorticity and current sheets that are highly correlated spatially, and  
 the sharp rolling-up of field lines in the current sheet.  At right, it  is shown that the Kelvin-Helmoltz instability of the current sheet (in a different sub-volume of the flow) can occur with different wavelengths.
} \end{center}  \label{f_pvr3}  \end{figure}

On the right of Fig. 8, a zoom in the current is shown, at a different location, with a small-scale wave-like structure in the sheet, this time with a characteristic scale rather like $\pi/50$, still well resolved numerically. These results confirm the instability, at high Reynolds number, of vorticity and current sheets that form  into rolls; it also shows, possibly for the first time in direct numerical simulations of MHD, a small-scale undulating structure likely linked to a Kelvin-Helmoltz instability, but at a slightly different scale than the rolls on the left. 
Such Kelvin-Helmoltz instabilities have been observed in the magnetosphere (see e.g. Hasegawa \etal 2004, Nykyri  \etal 2006) in a much more complex physical settings than what is being studied here, with compressibility, boundary and plasma kinetic effects in particular.
The occurence of a Kelvin-Helmoltz instability of current sheets has been widely studied in the presence of an imposed shear (see e.g. Knoll, 2002, 2003), but it has being obtained only recently numerically in three dimensions in a fully turbulent incompressible MHD flow.
Recent results using the Piecewise Parabolic Method for supersonic MHD turbulence also show rolled-up current sheets, in a series of studies performed in the context of modeling the interstellar medium (Kritsuk and Padoan, 2009).

\begin{figure} \begin{center}
\includegraphics[width=12.9cm, height=50mm]{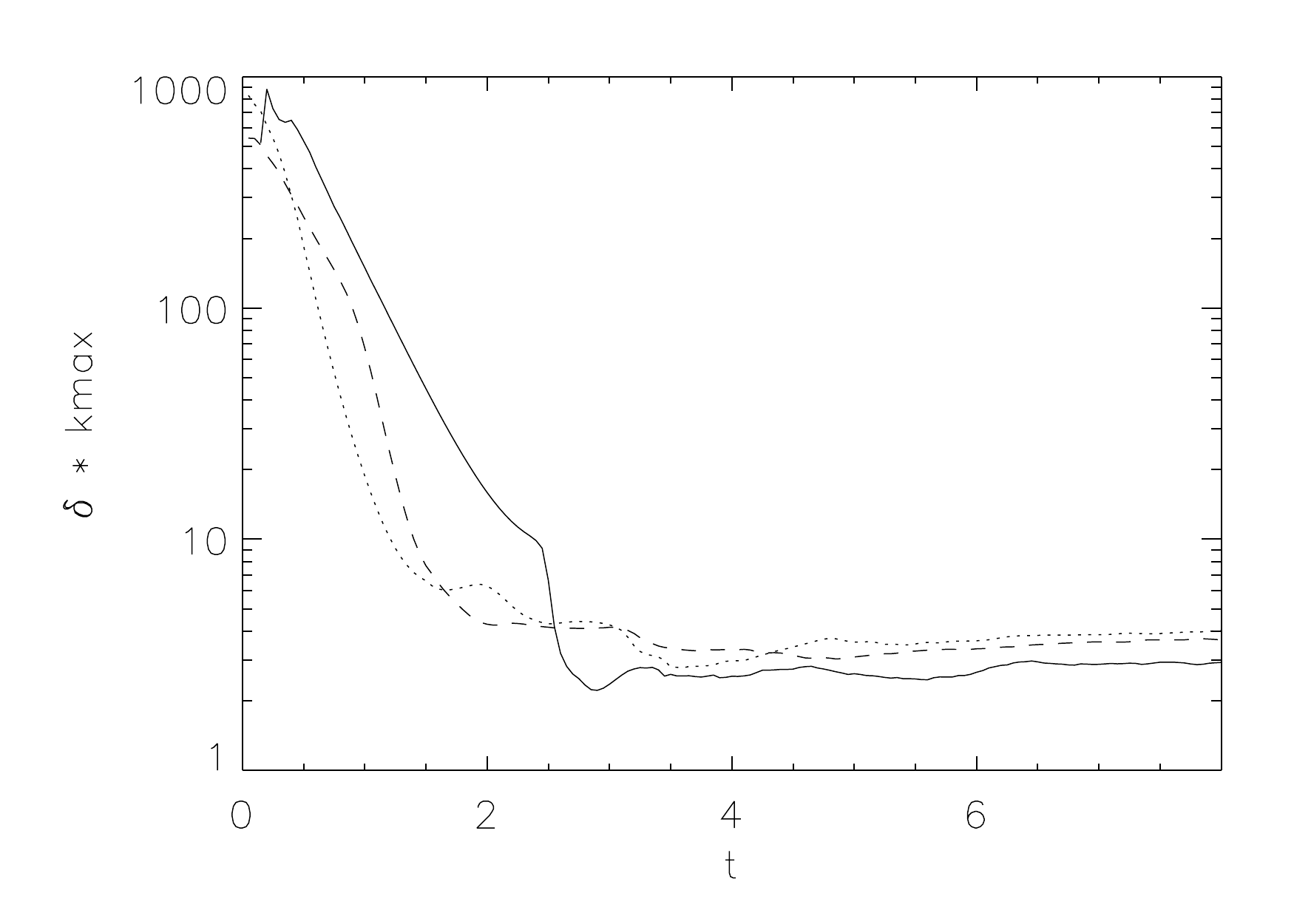}
\caption{ 
Temporal evolution of the logarithmic decrement $\delta$ (see Eq. \ref{eq:delta}) in units of maximal wavenumber $k_{max}$ for runs I6 (solid line), A6 (dash line) and C6 (dotted line), showing that each run is well resolved. See Table \ref{tab1} for the nomenclature of the runs.
The plot shows an abrupt event for $t\approx 2.5$ for the I6 flow (earlier for the magnetic skewness), linked to the quasi-collision of two current sheets.
} \label{f_temp_skew2}  \end{center} \end{figure}

One of the issue concerning the development of structures in a turbulent flow is to what degree the relevant vector fields are aligned; indeed, the nonlinearities in the MHD equations can be written in term of the Lamb vector ${\bf u} \times {\bf \omega}$, the Lorentz force ${\bf j} \times {\bf b}$ and Ohm's law ${\bf v} \times {\bf b}$. It can be shown that the primitive equations lead to an enhancement of the alignment between such vectors when they involve invariants such as the kinetic helicity (in the pure fluid case) or the cross-correlation between the velocity and magnetic field; such a local (as opposed to global) spatial alignment between the velocity and magnetic field is clearly observed both in the Solar Wind (using Ulysses data) and in DNS (Matthaeus \etal 2008, Servidio \etal 2008).
In the case of the flows studied in the present paper, alignment obtains again, but with also a peak in the PDF at orthogonality of vectors, a feature directly linked to the chosen initial conditions. However, a more detailed analysis reveals that the Lorentz force retains its full strength in all our TG runs (particularly so for the $B_A$ flows), in sharp contrast with the results of Servidio \etal 2008; moreover, the IMTG and $B_A$ flows both show a tendency toward Alfv\'enic alignment 
(${\bf v} \parallel {\bf b}$) while retaining a peak at orthogonality due to initial conditions (see Fig. \ref{f_full_symm_temp2}, left), whereas the $B_C$ flow displays a rather flat PDF, in contrast to the other two flows. Such alignment properties are not likely due to the enforcement of symmetries: Fig. \ref{f_full_symm_temp2} (left) gives the same data for the symmetric run (solid line) and the full DNS (circles) on a resolution of $512^3$ grid points, with no discernable discrepancy.

\subsection{Comparative behavior of the three initial conditions}

We have emphasized the properties of the IMTG flow until now. In this section, we want to examine briefly the development of small-scale turbulence for the other two types of initial conditions that we have written in \S \ref{s:eq} for a symmetric Taylor-Green MHD code;
Table \ref{tab1} summarizes several of the features of such flows, for the lowest to the highest Reynolds numbers.


\begin{table*} \caption{\label{tab1} 
Parameters of the runs described in this paper. 
The type of run (IMTG, $B_A$ or $B_C$ is indicated respectively as I, A or C, with run ``1'' or ``2'' for computations at a grid resolution of $128^3$ points, and runs labeled ``6''  on equivalent grids of $2048^3$ points. 
All flows have equal kinetic and magnetic energy $E_V=E_M=0.125$ at $t=0$; viscosity for run I1 is $2\times 10^{-3}$, for runs A2 and C2 , it is $10^{-3}$, and for runs I6, A6 and C6, it is equal to $6.25 \times 10^{-5}$, with in all cases $\nu=\eta$.
$R_V$ and $R_M^{\lambda}$ are respectively the Reynolds number at initial time and the magneticTaylor Reynolds number at the final time of the computation;
$\delta$ is the logarithmic decrement whose expression is given in Eq. (\ref{eq:delta}) and evaluated on the total energy spectrum; $k_{max}$ is the maximum wavenumber, with $N$ the linear number of points in the grid (here, 128 or 2048). 
 The ratio of magnetic to kinetic energy $E_M/E_V$ is the maximum value reached by this ratio as it evolves with time (excluding $t=0$).
The magnetic to kinetic modal ratio $\rho^E_x=E_M(k_x)/E_V(k_x)$ is
evaluated as an average over of the order of two turn-over times after the peak of dissipation, with $x=2,I$ and $\lambda$ for wavenumbers $k=2, \ k^I_M$ and $k^{\lambda}_M$ respectively, $k_M^{I,\lambda}$ being the magnetic integral and Taylor wavenumber (see Eq. \ref{eq:taylor} for the associated length scales).
Similarly, $\Omega_M/\Omega_V=<\j^2>/<\omega^2>$ is the maximum value attained by the ratio of magnetic to kinetic dissipation over time (recall that $\nu=\eta$).
Finally, $S_k^{max}$ and $S_k^{final}$ are the skewness evaluated on the velocity field at the time of maximum dissipation and the final time respectively.

$^{\ast}$ Note that in the case of the $B_C$ runs, the ratio of total magnetic to kinetic energies are given at their minimum value in time (this global ratio never goes above unity as time evolves). $^{\ast}$$^{\ast}$ For the $B_C$ enstrophies, the values are those at the first maximum after the initial time.
}  \vskip0.2truein
\halign{
#\hfil \quad & #\hfil \quad & #\hfil \quad & #\hfil \quad & 
#\hfil \quad & #\hfil \quad & #\hfil \quad & #\hfil \quad & 
#\hfil \quad & #\hfil \quad & #\hfil \quad        \cr  
%
  \multispan{11} \hrulefill \cr
\  \ \   Run                                    & $R_V$                        &$R_M^{\lambda}$ & $\delta k_{max}$       &
$E_M/E_V$                         & $\rho^E_2$               & $\rho^E_I$        & $\rho^E_{\lambda}$ & 
$\Omega_M/\Omega_V$   &$S_k^{max}$    &  $S_k^{final}$                     \cr 
%
  \multispan{11} \hrulefill \cr
   \noalign{\vskip4pt}
\ \ \ \ I1     &    800 & 110 &   1.9   & 
2.7   &  19   & 7  &   2    & 
5.7   &  1.5    & 1.15                \cr \noalign{\vskip3pt}   
\ \ \ \ I6     &    $2.3 \times 10^4$ & 2200 &   2.2   & 
6      &   3    & 28   &   18   & 
6.1   &   2.8    & 1.15               
                \cr \noalign{\vskip3pt}   
%
  \multispan{11} \hrulefill \cr
\ \ \ \ A2     &    1200 & 160 &   2.1   &                        
1.5   &  0.3   & 1.9  &   1.8     & 
2.2   &   2.2    & 0.6                \cr \noalign{\vskip3pt}
\ \ \ \ A6     &    $2\times 10^4$ & 900 &   3   &              
1.9   &  0.4   & 2.2  &   1.6     & 
4.4   &   2.2    & 0.8               
                \cr \noalign{\vskip3pt}   
%
  \multispan{11} \hrulefill \cr
\ \ \ \ C2     &    1200 & 100 &   0.8   &             
0.35 \ $^{\ast}$   &  0.09   & 1.2  &   1.7     & 
2. $^{\ast}$$^{\ast}$  &   2.1    & 0.55                \cr \noalign{\vskip3pt}
\ \ \ \ C6     &    $2\times 10^4$ & 900 &   2.8   &            
0.65 \ $^{\ast}$   &  0.06   & 2.2  &   1.5     & 
3.  $^{\ast}$$^{\ast}$ &   2.8    & 1.6                \cr \noalign  {\vskip3pt}
%
  \multispan{11} \hrulefill \cr
 }  \end{table*}

One possible way to monitor the development of small scales is to examine the temporal development of the
 logarithmic decrement. The isotropic total energy spectrum $E_T(k,t)$ defined in the usual manner by averaging in spherical shells of width $\Delta k=1$ can be fitted with the following functional form (Sulem \etal 1983):
\begin{equation}
E_T(k,t) = \frac{1}{2} \int_{|\bm k|}^{|\bm k|+1} ([v(\bm k,t)]^2 + [b(\bm k,t)]^2) \ d^3k
       =  C(t) k^{-n(t)} e^{-2\delta(t)k} \ .
\label{eq:delta} \end{equation}
 The so-called logarithmic decrement $\delta$ is a measure of the smallest scale reached by the flow at a given time and can be compared to the grid resolution $3/N$ (taking aliasing into account); this criterion thus allows for measuring the accuracy of the computation at any given time, comparing $\delta$ and $3/N$. The analysis of the temporal evolution of $\delta$ was performed in the ideal (non dissipative) case both in two dimensions (Frisch \etal 1983) 
 and in three dimensions (Lee \etal 2008). In the latter case,
 it is found to decay exponentially ($\delta(t) = \delta_0 e^{-t/\tau_i}$),
with two different values of $\tau_i$ corresponding to two different ideal instabilities: first, a classical thinning of current and vorticity sheets occurs, due to the large-scale effect of shear; this phase is followed by a faster near-collision of two near-by current sheets, due to a favorable magnetic pressure gradient associated with the Lorentz force. The computation in the ideal case has to be stopped once $\delta\sim 3/N$; but in the dissipative case, the run can be continued for long times provided the viscous and resistive terms are strong enough to arrest the process of small-scale formation.

\begin{figure} \begin{center}
\includegraphics[width=8.5cm, height=50mm]{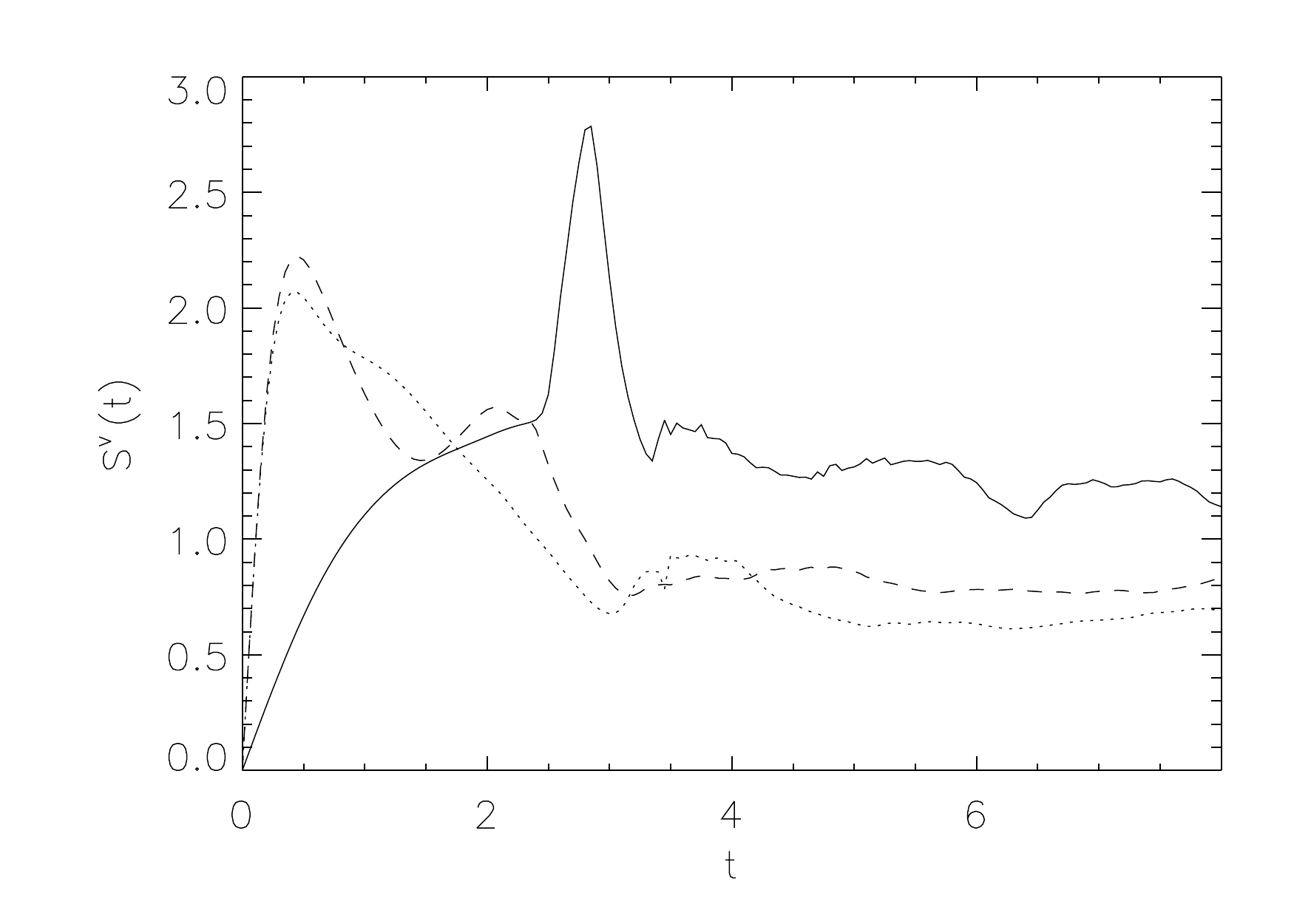}
\includegraphics[width=8.5cm, height=50mm]{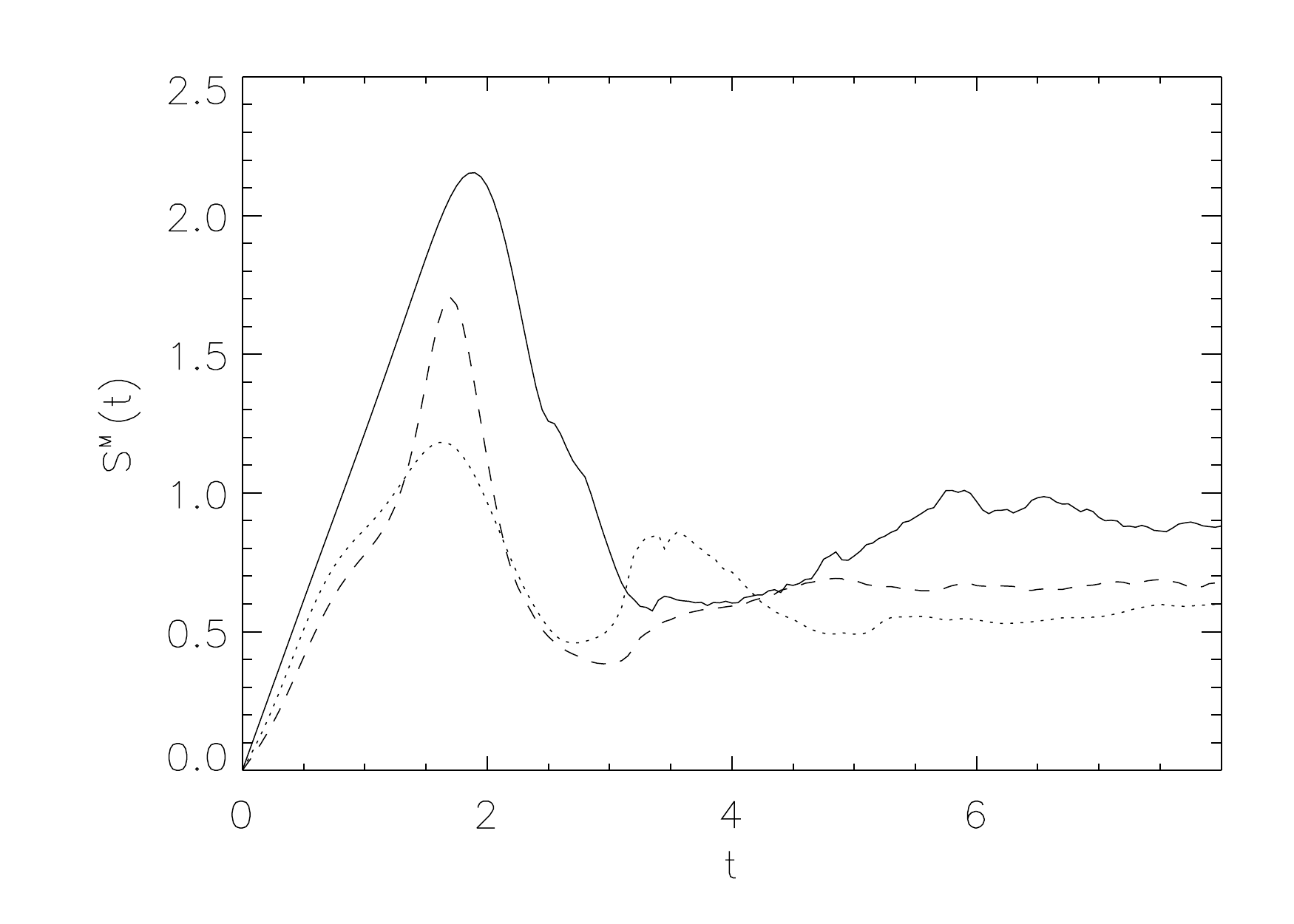}
\caption{ 
{\it Left:} Temporal evolution of the skewness, i.e. the normalized third-order moment of the velocity field, all flows computed on an equivalent grid of $2048^3$ points, for runs I6 (solid line), A6 (dash line) and C6 (dotted line) (see Table \ref{tab1}). 
{\it Right:} Skewness built on the magnetic field for the same three runs.
Both plots show an abrupt event for $t\approx 2.5$ for the IMTG flow (earlier for the magnetic skewness), linked to the near-collision of two current sheets leading to a quasi rotational discontinuity.
} \label{f_temp_skew3}  \end{center} \end{figure}

In this context, we now examine the temporal behavior of the logarithmic decrement in the dissipative case (see Fig. \ref{f_temp_skew2}). We observe that at the same time as for the ideal case, an abrupt change occurs for the IMTG flow (solid line), followed by a saturation of the decreasing of $\delta$ when dissipation sets in, at a wavenumber wich, in the fluid case, varies as $R_V^{3/4}$. Only the IMTG flow displays a substantial acceleration of the development of small scales, also discernible in the evolution of the normalized third order moment of the velocity and magnetic field derivatives, but the three flows are well resolved at all times since $\delta k_{max} \ge 2$. The development of small scales appears more rapid for the $B_{A,C}$ flows, since they reach their plateau in $\delta$ at an earlier time ($\approx 2$) than for the IMTG flow. Similar behavior can be observed on the temporal evolution of the kinetic and magnetic skewness (normalized third-order moment of the velocity and magnetic field), as seen in Fig. \ref{f_temp_skew3}. Note that these moments systematically reach higher values than in the pure fluid case, and that one finds for all three flows higher values of the velocity skewness compared to its magnetic counterpart.

The three types of runs studied here have almost identical initial conditions, from an ideal point of view: same energy, same equipartition at $t=0$ between kinetic and magnetic energy, same zero magnetic helicity, same weak correlation between the velocity and the magnetic field (0 to 4\% in relative terms).
They also have the same type of structures in the current and vorticity, with sheets and rolls. However, they differ in their dynamics.
For example, for the IMTG runs, the magnetic to kinetic energy and enstrophy ratios are all larger than unity, particularly so at $k=2$ (and
the length scales built on the velocity are smaller than those build on the magnetic field).  For the $B_A$ runs, the kinetic and magnetic variables are close to being in balance whereas in the $B_C$ case, it is now the kinetic quantities that dominate. The strongest differences between the three flows is the dynamics of the low-k modes, whereas, for all runs, there is an excess of magnetic energy for $k\ge k^M_I$, with $k^M_I$ the magnetic integral wavenumber (see Table \ref{tab1}). These issues will be examined further in the future (Lee \etal 2009). 


\section{Discussion and conclusion} \label{s:conclu}

In this paper, we have implemented numerically the symmetries of the Taylor-Green flow generalized to MHD and have thus been able to study the dynamics of MHD turbulence at higher Reynolds number than what can be reached in a full DNS computation. We checked that no spurious error develops when using such codes, and we have explored the properties of three types of initial conditions, stressing the evolution in one particular case, the IMTG flow. We show that this flow develops an energy spectrum, $E_T(k)\sim k^{-2}$ which is in agreement with the prediction of weak MHD turbulence, the spectrum being obtained as well for 
$E_T(k_{\perp})\sim k_{\perp}^{-2}$.
The detailed analysis of the other two sets of initial conditions that follow the symmetries of the Taylor-Green flow are left for future work (Lee \etal 2009). 

In the study of MHD turbulence and of the dynamo problem, the Taylor-Green flow has proved quite useful. Other methods to progress in our understanding of MHD turbulence is to resort to modeling. Andrew Soward (1972) pioneered this in developing an Eulerian-Lagrangian approach to the kinematic dynamo (see also Soward and Roberts, 2008, for a recent analysis). Other models can be derived and used (see e.g., Kraichnan and Nagarajan, 1967, Pouquet \etal 1976, Yoshizawa 1990, Holm 2002a, 2002b, Graham \etal 2009), including numerical (see e.g. Meneguzzi \etal 1996). Using both  the implementation of symmetries and such models may prove a fruitful approach to explore parameter space (Pouquet \etal 2009).
For example, it has been known for a long time that the amount of correlation between the velocity and the magnetic field modifies the spectral indices of the Els\"asser variables (and hence of the other energy spectra), with a steeper slope for the dominant mode (see Grappin \etal 1983, in the context of isotropic turbulent closures, Galtier \etal 2000, for a weak turbulence anisotropic theory, Politano \etal 1989, for the numerical confirmation of such a model in the two-dimensional isotropic case, and for the three-dimensional case, Politano \etal 1995). 
A small amount of correlation alters the energy spectrum only slightly, making data analysis difficult. 

 Among the other parameters of potential importance in the study of magnetohydrodynamic turbulence is the magnetic Prandtl number $\nu/\eta$, 
the presence of sizable global (as opposed to local) correlation between the velocity and the magnetic field, 
the ratio of magnetic to kinetic energy, and the effect of an externally imposed uniform magnetic field. Compressibility, rotation, stratification are other external agents of imposed anisotropies, and important for plasma experiments, the presence of small-scale kinetic effects such as a Hall current or pressure anisotropies may also alter the dynamics of MHD flows. Using imposed symmetries to enhance the numerically attainable Reynolds number at a given cost may be a promising venue in many of these cases.
 
 \vskip0.5truein
{\it Acknowledgments}

{\sl It is our pleasure to dedicate this paper to Andy Soward in whose honor the Nice meeting was held.}
Computer time was provided through NSF MRI -CNS-0421498, 0420873 and 0420985, NSF sponsorship of NCAR, the University of Colorado, and a grant from the IBM Shared University Research (SUR) program.
Ed Lee was supported in part from NSF IGERT Fellowship in the Joint Program in Applied
Mathematics and Earth and Environmental Science at Columbia.
Visualizations use the VAPOR software for interactive visualization and analysis of terascale datasets (Clyne \etal 2007)

\end{document}